\documentclass[fleqn,usenatbib,usedcolumn]{mnras}
\usepackage[british]{babel}             
\usepackage{newtxtext}                  
\usepackage{newtxmath}                   
\let\la=\lesssim 
\let\ga=\gtrsim
\usepackage[T1]{fontenc}                
\usepackage{bm}
\usepackage[pdftex]{graphicx}
\usepackage{times}
\usepackage{url}

\newcommand{\pvd}{\mbox{$\sigma_{12}$}}

\newcommand{\hMpc}{\mbox{\,$h^{-1}$\,Mpc}}
\newcommand{\hMpcinv}{\mbox{\,$h$\,Mpc$^{-1}$}}
\newcommand{\kms}{\mbox{$\mbox{\,km\,s}^{-1}$}}

\newcommand{\vbar}{\mbox{$\overline{v}_{12}$}}

\title[GAMA: Pairwise velocity dispersion]{Galaxy and Mass Assembly (GAMA): Small--scale anisotropic galaxy clustering and the pairwise velocity dispersion of galaxies}

\author[J.~Loveday et al.]
{{\parbox{\textwidth}{\raggedright J.~Loveday,$^{1}$\thanks{E-mail:~J.Loveday@sussex.ac.uk}
L.~Christodoulou,$^{1}$
P.~Norberg,$^{2}$
J.A.~Peacock,$^{3}$
I.K.~Baldry,$^{4}$
J.~Bland-Hawthorn,$^{5}$
M.J.I.~Brown,$^{6}$
M.~Colless,$^{7}$
S.P.~Driver,$^{8,9}$
B.W.~Holwerda,$^{10}$
A.M.~Hopkins,$^{11}$
P.R.~Kafle,$^{8}$
J.~Liske,$^{12}$
A.R.~Lopez-Sanchez,$^{11,13}$
E.N.~Taylor$^{14}$
}}\\
\vspace{0.4cm}\\
{\parbox{\textwidth}{\raggedright 
$^{1}$Astronomy Centre, University of Sussex, Falmer, Brighton BN1 9QH, UK
\\
$^{2}$ICC \& CEA, Department of Physics,
Durham University, South Road, Durham DH1 3LE, UK
\\
$^{3}$Institute for Astronomy, University of Edinburgh, Royal
Observatory, Blackford Hill, Edinburgh EH9 3HJ, UK
\\
$^{4}$Astrophysics Research Institute, Liverpool John Moores University, IC2, Liverpool Science Park, 146 Brownlow Hill, Liverpool, L3 5RF, UK
\\
$^{5}$Sydney Institute for Astronomy, School of Physics, University of
Sydney, NSW 2006, Australia
\\
$^{6}$School of Physics, Monash University, Clayton, Victoria 3800, Australia
\\
$^{7}$Research School of Astronomy \& Astrophysics,
The Australian National University,
Cotter Road,
Weston Creek, ACT 2611,
Australia
\\
$^{8}$International Centre for Radio Astronomy Research (ICRAR),
    The University of Western Australia, 35 Stirling Highway, Crawley,
    WA6009, Australia
\\
$^{9}$School of Physics \& Astronomy, University of St Andrews, North
    Haugh, St Andrews, KY16 9SS, UK
\\
$^{10}$Department of Physics and Astronomy, University of Louisville, Louisville, KY 40292, USA
\\
$^{11}$Australian Astronomical Observatory, PO Box 915, North Ryde, NSW 1670, Australia
\\
$^{12}$European Southern Observatory, Karl-Schwarzschild-Str.~2, 85748
Garching, Germany
\\
$^{13}$Department of Physics and Astronomy, Macquarie University, NSW 2109, Australia
\\
$^{14}$Centre for Astrophysics \& Supercomputing, Swinburne University of Technology, P.O. Box 218, Hawthorn, VIC 3122, Australia
\\
}}}

\date{Accepted XXX. Received YYY; in original form ZZZ}

\pubyear{2017}

\begin{document}
\label{firstpage}
\pagerange{\pageref{firstpage}--\pageref{lastpage}}
\maketitle

\begin{abstract}
  The galaxy pairwise velocity dispersion (PVD) can provide important tests of
  non-standard gravity and galaxy formation models.
  We describe measurements of the PVD of galaxies in the
  Galaxy and Mass Assembly (GAMA) survey as a
  function of projected separation and galaxy luminosity.
  Due to the faint magnitude limit ($r < 19.8$) and highly-complete 
  spectroscopic sampling of the GAMA survey, 
  we are able to reliably measure the PVD
  to smaller scales ($r_\bot = 0.01 \hMpc$) than previous work.
  The measured PVD at projected separations $r_\bot \la 1 \hMpc$
  increases near-monotonically with increasing luminosity from
  $\pvd \approx 200 \kms$ at $M_r = -17$ mag
  to $\pvd \approx 600 \kms$ at $M_r \approx -22$ mag.
  Analysis of the \citet{Gonzalez-Perez2014} {\sc galform} semi-analytic model
  yields no such trend
  of PVD with luminosity: the model over-predicts the PVD for faint galaxies.
  This is most likely a result of the model placing too many low-luminosity
  galaxies in massive halos.
\end{abstract}


\section{Introduction} \label{sec:intro}

The pairwise velocity dispersion (PVD, \pvd), 
the dispersion in relative peculiar velocity of galaxy pairs,
has an illustrious history in observational cosmology.
It was first measured in 1973 by \citet{Geller1973},
and soon became popular as a way of estimating the mean mass density
of the Universe, $\Omega_m$, via the cosmic virial theorem or
cosmic energy equation
\citep[e.g.][]{Peebles1976a,Peebles1976,Peebles1979,Bean1983,Davis1983,Bartlett1996}.
In fact, these measurements provided perhaps the first evidence that we live in
a Universe which has a sub-critical mass density, $\Omega_m < 1$.

Use of the PVD to constrain cosmological parameters then fell out of favour,
largely due to its sensitivity to the presence or absence of rich clusters
in the survey data used \citep{Mo1993}.

Nevertheless, knowledge of the (non-linear) PVD is required when modelling
the linear, large-scale \citet{Kaiser1987} infall in order to constrain
the growth rate of structure (e.g. \citealt{Peacock2001,Guzzo2008,Blake2013}). 
The PVD is an important quantity
for modeling the galaxy redshift--space correlation function,
and can be used to test predictions of galaxy formation and evolution models,
the focus of this paper, and of the cold dark matter paradigm in general. 

Recently, interest in use of the PVD as a cosmological diagnostic
has been reawakened, both due to the availability of large spectroscopic surveys
which encompass fair samples of the Universe,
and due to theoretical developments in modified gravity.
There have been several recent efforts to model modified gravity
using $N$-body simulations,
allowing one to compare the predictions of small-scale
galaxy dynamics, e.g. \citet{Fontanot2013,Falck2015,Winther2015,Bibiano2017}. 
In particular, \citet{Hellwing2014} have shown that the PVD provides
one of the most
sensitive diagnostics of modified gravity, with some of these models
predicting dispersions about 30 per cent larger or smaller
than General Relativity.

Since the first measurements of \citet{Geller1973},
the PVD has been measured for most redshift surveys,
e.g. \citet{Bean1983,Davis1983,Loveday1996,Jing1998,Jing2001,Landy2002,Zehavi2002,Hawkins2003,Jing2004,Li2006b,VanDenBosch2007,Cabre2009}.
Predictions of the PVD from HOD models and/or simulations have been made by 
\citet{Slosar2006,Li2007,Tinker2007,VanDenBosch2007}.
A good summary of previous results for the overall PVD, 
i.e. measured over all galaxy types and a wide range of scales,
is provided by \citet{Landy2002}.
Most estimates range from around 300 to 600 \kms, and are 
sensitive to the presence or absence of rich clusters in the data used
\citep{Mo1993}.
The advent of large redshift surveys, such as the two-degree Field 
Galaxy Redshift Survey \citep[2dFGRS;][]{Colless2001}
and the Sloan Digital Sky Survey \citep[SDSS;][]{York2000},
enabled detailed studies of the dependence of the PVD on galaxy type and scale.
\citet{Hawkins2003} measure the PVD for 2dFGRS galaxies, finding a peak
$\pvd \approx 600 \kms$ at projected separations
$r_\bot \approx 0.2$--0.8 \hMpc, with \pvd\ declining to 300--400 \kms\
at smaller and larger scales, consistent with contemporary semi-analytic model
predictions.
\citet{Jing2004} find that at a scale of $k = 1 \hMpcinv$,
the 2dFGRS PVD has a minimum value of  $\pvd \approx 400 \kms$ for 
galaxies of luminosity $M^* - 1$, 
increasing rapidly for both fainter and brighter galaxies.
\citet{Li2006b} measure the PVD for SDSS galaxies as a function of 
luminosity and stellar mass as well as other galaxy properties.
Consistent with \citet{Jing2004}, 
they find that the PVD measured at $k = 1 \hMpcinv$ has a minimum value of 
$\pvd \approx 500 \kms$ for galaxy luminosities around $M^* - 1$, 
increasing somewhat for less luminous galaxies, and markedly 
(to $\pvd \approx 700 \kms$) for the most luminous galaxies in the sample.
They also find that red galaxies have systematically higher PVDs 
than blue galaxies, particularly for less luminous galaxies.
In a followup paper, \citet{Li2007} compare the clustering and PVD of
SDSS galaxies with semi-analytic models,
finding that the models over-predict the clustering strength and PVD
for sub-$L^*$ galaxies, particularly at small scales.

The Galaxy and Mass Assembly (GAMA) survey \citep{Driver2011} provides an
ideal opportunity for a new measurement of the PVD due to 
(i) being two magnitudes fainter than the SDSS main galaxy sample, 
and (ii) having very high ($> 98$ per cent)
spectroscopic completeness, even in high-density regions.
The latter point means that completeness corrections for ``fibre collisions''
are not an issue with GAMA data.
We utilise the three equatorial regions in  GAMA-II \citep{Liske2015},
covering a total area of 180 square degrees, and including galaxies down to
Petrosian $r$-band apparent magnitude $r = 19.8$.
The GAMA-II database has previously been used to measure the projected
galaxy clustering in bins of stellar mass and luminosity \citep{Farrow2015}
and to measure the growth rate of large-scale structure via linear-regime
redshift-space distortions (RSD) \citep{Blake2013}.
Here, we focus on measuring RSD in the non-linear regime,
$r_\bot \la 10 \hMpc$.

The paper is structured as follows.
We discuss the GAMA data, mock and random catalogues in 
Section~\ref{sec:data} and
measurement of two-dimensional and projected correlation functions 
in Section~\ref{sec:xi2d}.
In Section~\ref{sec:dist}, we describe two models for the redshift-space
correlation function, and demonstrate that the pairwise velocity distribution
function is close to exponential.
We test three different ways of measuring the PVD
using mock catalogues in Section~\ref{sec:method}.
PVDs for the GAMA data in luminosity bins, along with a comparison of mock
predictions, are shown in Section~\ref{sec:results};
we conclude in Section~\ref{sec:concs}.
Throughout, we assume a Hubble constant of $H_0 = 100 h$ km~s$^{-1}$~Mpc$^{-1}$ 
and an $\Omega_M = 0.25$, $\Omega_\Lambda = 0.75$ cosmology in
calculating distances, co-moving volumes and luminosities.
Uncertainties on all results from GAMA data and mocks are based on
jackknife sampling and from the scatter between realisations, respectively.

\section{Data, mock and random catalogues} \label{sec:data}

\subsection{GAMA data}

Our observed sample consists of galaxies from the GAMA-II equatorial regions
G09, G12 and G15, each $5 \times 12$ degrees in extent and 98 per cent
spectroscopically complete to $r = 19.8$ mag \citep{Liske2015}. 
Specifically, galaxy coordinates and magnitudes come from TilingCatv46 
\citep{Baldry2010}.
Redshifts, corrected by the multiattractor flow model of \cite{Tonry2000},
as described by \citet{Baldry2012}, are taken from DistancesFramesv14.
$K$-corrections to reference redshift $z_0 = 0.1$ \citep{Blanton2007}
and 4th-order polynomial fits are obtained from kCorrectionsv05 
\citep{Loveday2012}.

In order to estimate errors on our results, we subdivide each GAMA field
into three $4 \times 5$ deg regions, and determine the covariance by
omitting each of the nine jackknife regions in turn.
The median velocity uncertainty in GAMA is $33 \kms$ \citep{Baldry2014}, 
significantly less than the smallest measured velocity dispersions, 
and so we quote PVDs uncorrected for these measurement errors.
Similarly, we ignore the effect of blended galaxy spectra,
where galaxies are either lensed or overlapping \citep{Holwerda2015},
since this affects only 0.05 percent of the GAMA sample.

\subsection{{\sc galform} mock catalogues}

We compare our GAMA results with mock galaxy catalogues based on the 
Millennium-WMAP7 Simulation \citep{Guo2013a} and the 
\citet{Gonzalez-Perez2014} {\sc galform} model, 
with lightcones produced using the method of \citet{Merson2013};
see \citet{Farrow2015} for further details of these GAMA mocks.
Specifically, we queried the table \verb|GAMA_v1..LC_multi_Gonzalez2014a|
via the Durham-hosted Virgo--Millennium 
Database\footnote{Millennium DB at \url{http://virgodb.dur.ac.uk}}
\citep{Lemson2006}.

We utilize 26 mock realisations of the three equatorial 
GAMA fields (G09, G12 and G15), selecting galaxies down to apparent 
SDSS $r$-band magnitude $r < 19.8$ mag.
These mocks were extracted from the Millennium-WMAP7 simulation cube
using random observer position and orientations.
As such they are not independent,
but do allow some assessment of sample variance.
Since the mocks provide both an observed and a cosmological redshift, 
we can make a direct estimate of the PVD to compare with our
clustering-based PVD estimators.
Covariance estimates for the mocks come from comparing the 26 realisations.
In practice, we use only the diagonal elements of the covariance matrix.
Since we focus on the small-scale PVD, the relatively small number and
lack of true independence of the mocks is not a serious issue.

\subsection{Random catalogues}

In order to account for the survey boundaries and selection effects,
we generate random catalogues obeying the same mask and selection function
as the GAMA data, but without clustering.
The mask and selection function are derived independently from those of 
\citet{Farrow2015}; we have checked that we obtain consistent results
for the projected correlation function (see Appendix~\ref{sec:farrow_comp}).
The mock catalogues have a simple mask corresponding to the RA--dec 
boundaries of the GAMA equatorial regions.
The radial distribution of random points for analysing the mocks is obtained
by taking the mock galaxy redshifts from all 26 realisations of
the three GAMA fields;
large-scale structure in individual realisations is rendered invisible in the
combined distribution.
The following subsections describe the survey mask and radial selection function
for the GAMA data.

\subsubsection{Survey mask}

\begin{figure}
\includegraphics[width=\linewidth]{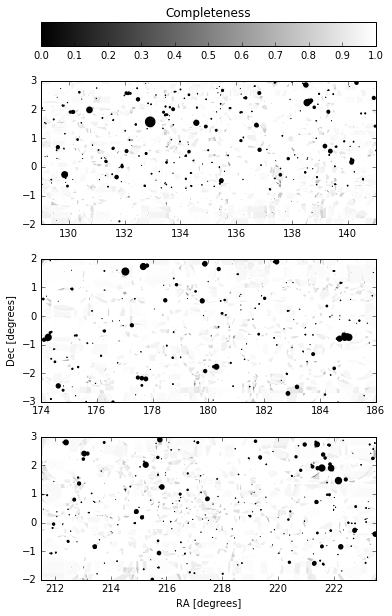}
\caption{Spectroscopic completeness mask for the GAMA-II regions.
Black regions correspond to holes cut around bright stars and 
SDSS imaging defects.}
\label{fig:mask}
\end{figure}

Since GAMA-II target selection was made from SDSS DR7 $r$-band imaging,
we mask out regions of $r$-band imaging identified by the SDSS photometric 
pipeline as any of {\sc bleeding, bright\_star, trail, hole}\footnote{See 
\url{http://www.sdss.org/dr7/algorithms/masks.html}}.
In addition we mask out areas around bright stars ($V < 12$ mag)
in the Tycho and Hipparcos catalogues --- see \citet{Baldry2010}
for details.

In order to map spectroscopic completeness as a function of position on the sky,
we obtain a list of GAMA 2dF field centres from the table AATFieldsv25.
The (zero-weight) mask regions and (unit-weight) spectroscopic fields
are then combined using the {\sc pixelize},  {\sc snap} and  {\sc balkanize}
commands in  {\sc mangle} \citep{Hamilton2004,Swanson2008}.
The result is a list of polygons defined by overlap regions of the 2dF fields
with masked regions set to zero weight.
We then set the weight of each non-masked polygon to its 
spectroscopic completeness 
by dividing the number of main-survey targets (survey\_class $> 3$)
with reliable redshifts ($nQ > 2$) by
the number of targets within each polygon.
Finally, we trim the polygons to lie within the equatorial coordinate ranges
of the three GAMA regions, namely
$\alpha = {(129.0, 141.0), (174.0, 186.0), (211.5, 223.5)}$,
$\delta = {(-2.0, 3.0), (-3.0, 2.0), (-2.0, 3.0)}$, 
for G09, G12 and G15 respectively.
The resulting spectroscopic completeness maps are shown in Fig.~\ref{fig:mask}.
Angular coordinates of random points are generated using the  
{\sc mangle ransack} 
command with density proportional to the completeness within each polygon.

\subsubsection{Radial selection function}

When analysing samples that are not volume-limited,
the radial coordinates of random points are generated from each sample using the
joint stepwise maximum likelihood (JSWML) method of \citet{Cole2011},
as adapted for use with GAMA by \citet{Loveday2015},
assuming evolution parameters $P = Q = 1$.

For volume-limited samples, 
we distribute points drawn at random from a distribution uniform in 
comoving volume modulated by the density-evolution factor $10^{0.4 P z}$
\citep[][equation 5]{Loveday2015}.
Limiting redshifts for each volume-limited sample are chosen such that the
distribution of individual $K$-corrections for galaxies close to
the limiting redshift results in a sample that is 95 percent complete.

\subsubsection{Comparison with previous GAMA clustering measurements}

Since both the angular mask and radial selection function for the random 
catalogues have been derived independently from a previous measurement
of galaxy clustering from the GAMA data \citep{Farrow2015},
we compare our clustering estimates for a number of galaxy subsamples in
Appendix~\ref{sec:farrow_comp}.
We find results that are in excellent agreement on all small scales.

\subsection{Data subsamples} \label{sec:subsamples}

\begin{table}
 \caption{GAMA and mock galaxy subsamples.
 GAMA magnitudes are evolution-corrected $^{0.1}M_r$; 
 mock magnitudes, $^{0.0}M_r$, are taken directly from the mock catalogues.
}
 \label{tab:samples}
 \begin{math}
 \begin{array}{lrrrc}
 \hline
\mbox{Name} & \mbox{GAMA mag} & N_{\rm gal} & \mbox{Mock mag} & N_{\rm mock}\\
\hline
{\rm V0} & [-23, -20] &  41757 & [-23, -20] & 53878 \pm 3038 \\
{\rm M1} & [-23, -22] &  3730 & [-23.47, -22.36] & 3507 \pm 108 \\
{\rm M2} & [-22, -21] &  37904 & [-22.36, -21.02] & 38278 \pm 934 \\
{\rm M3} & [-21, -20] &  68791 & [-21.02, -19.85] & 67704 \pm 2468 \\
{\rm M4} & [-20, -19] &  43105 & [-19.85, -18.19] & 44115 \pm 2133 \\
{\rm M5} & [-19, -18] &  17550 & [-18.19, -17.00] & 17224 \pm 1570 \\
{\rm M6} & [-18, -17] &  6037  \\
{\rm M7} & [-17, -16] &  2080  \\
{\rm M8} & [-16, -15] &  805  \\
 \hline
 \end{array}
 \end{math}
\end{table}

We measure the PVD of GAMA galaxies in bins of absolute magnitude,
as summarized in Table~\ref{tab:samples}.
For all GAMA samples, we employ individual $K$-corrections to rest-frame
$z = 0.1$ and assume luminosity
evolution given by $Q=1$ as described in \citet{Loveday2015}.
We use a superscript prefix of 0.1, $^{0.1}M_r$, to indicate an
absolute magnitude $K$-corrected to a passband blueshifted by $z = 0.1$.
This is done for the GAMA data to allow comparison with the results of
\citet{Li2006b}.
A superscript prefix of 0.0 indicates an absolute magnitude $K$-corrected to
the rest-frame of the galaxy, as appropriate for the mock data.

Because the mocks do not provide an exact match to GAMA in terms of
luminosity function, $K$-corrections and evolution, 
we cannot obtain a fair comparison by using the same magnitude limits.
Instead, we select samples matched on number density, as is common in the
literature, e.g. \citet{Berlind2003,Zheng2005,Contreras2013,Farrow2015}.

Allowing for the fact that the GAMA mask removes about 0.7 percent of
the survey area \citep{Baldry2010} and that the GAMA redshift incompleteness 
is about 1.5 percent \citep{Liske2015},
one would expect the GAMA catalogues to contain about 2.5 percent fewer
galaxies than the mocks if they have the same underlying number 
density\footnote{Although the GAMA regions are underdense
with respect to SDSS by about 15 percent within $z < 0.1$ \citep{Driver2011},
there is no evidence that this underdensity extends out to larger redshifts.
The overall sample variance of GAMA is expected to be about 3 percent
\citep{Driver2010}.}.
Starting with an absolute magnitude threshold of $^{0.1}M_r = -23$ mag, 
we count the number of GAMA galaxies brighter than this threshold within
redshift $z < 0.65$.
We find the corresponding mock absolute magnitude threshold 
that gives 1.025 times as many galaxies when averaged 
over the 26 mock realisations within the same redshift limit.
This process is repeated for the remaining magnitude bins;
the corresponding magnitude limits for the mocks are given in 
Table~\ref{tab:samples}.
Note that the GAMA magnitudes are $K$- and evolution-corrected to $z_0 = 0.1$,
assuming luminosity evolution given by $Q=1$,
whereas we use absolute $r$-band magnitudes (\verb|SDSS_r_rest_abs|)
taken directly from the mock catalogue without any evolution correction.
For sample M5, we set the mock faint magnitude limit to be $-17$ mag,
even though this sample contains fewer galaxies than the corresponding 
GAMA sample.
This is due to the resolution limit of the Millennium Simulation:
samples fainter than $^{0.0}M_r \approx -17$ mag will be incomplete
in a halo-dependent way.
This incompleteness in the mock catalogues may explain the
spuriously high clustering signal measured by \citet[][Fig.~11]{Farrow2015}
for the $-18 <\ ^{0.0}M_r < -17$ mag mock sample.

For testing our methods in Sections \ref{sec:dist} and \ref{sec:method}, 
we employ a volume-limited sample (V0), 
with $^{0.1}M_r < -20$ mag and redshift $z < 0.258$.
These limits are chosen in order to roughly maximize the number of galaxies
in a volume-limited sample.
We choose the same limits for the mocks, 
since number densities are very similar at this magnitude.
While the corresponding redshift limit will not be identical for the mock
catalogues, this is not an issue, as the random distribution for analysing
mock galaxy clustering is generated from the mocks themselves.

The real-space correlation function for the V0 mock sample is
well-fitted on scales
$r \la 16 \hMpc$ by a power-law with $\gamma = 1.81$, $r_0 = 5.6 \hMpc$.
From numerical integration of this power-law,
we find that the variance of galaxy counts
in $8 \hMpc$ radius spheres is very close to 1, $\sigma^2_{8, g} \approx 0.98$.
Since the simulations assume the WMAP7 cosmology ($\Omega_m = 0.272$,
$\sigma_8 = 0.807$, \citealt{Guo2013a}), the bias of this mock galaxy
sample is $b = \sigma_{8, g}/\sigma_8 \approx 1.23$ and hence the expected
value of the redshift space distortion parameter is given by
$\beta = \Omega_m^{0.6}/b \approx 0.37$.

Our primary results (Section~\ref{sec:results}) show galaxies in 
non-volume-limited bins of luminosity (M1--M8), as well as five samples
drawn from a single volume-limited sample.

\section{Measuring the correlation function} \label{sec:xi2d}

Our measurements of the PVD are based on the two-dimensional 
galaxy correlation function $\xi(r_\bot, r_\|)$;
the excess probability above random of finding two galaxies separated by $r_\|$
along the line of sight (LOS) and $r_\bot$ perpendicular to the LOS.
These separations are calculated in the usual way \citep[e.g.][]{Fisher1994}.
Two galaxies with position vectors ${\bm r}_1$ and ${\bm r}_2$
are separated by vector ${\bm s} = {\bm r}_2 - {\bm r}_1$.
For an observer at the origin, the vector to the midpoint of the pair 
is given by 
${\bm l} = ({\bm r}_1 + {\bm r}_2)/2$.
The LOS and perpendicular separations of the galaxies 
are then given by
$r_\| = |{\bm s}.\hat{{\bm l}}|$, with $\hat{{\bm l}}$ being
the unit vector in the direction of ${\bm l}$, and
$r_\bot = \sqrt{{\bm s}.{\bm s} - r_\|^2}$.

To estimate $\xi(r_\bot, r_\|)$, we use the \citet{Landy1993} estimator,
\begin{equation}
  \label{eq:ls}
  \xi(r_\bot, r_\|) = \frac{DD - 2DR + RR}{RR},
\end{equation}
where $DD$, $DR$ and $RR$ are the normalised and weighted numbers of data-data,
data-random and random-random pairs in a given $(r_\bot, r_\|)$ bin.
The random points are generated as described in the previous section.
For non-volume-limited samples, 
the pair counts are weighted to allow for the declining selection function
with redshift, giving a minimum-variance estimator
\citep{Hamilton1993}.
Each galaxy pair is given a weight
\begin{equation}
w_{ij} = \{[1 + 4 \pi \overline{n}(z_i) J_3(s_{ij})]
[1 + 4 \pi \overline{n}(z_j) J_3(s_{ij})]\}^{-1},
\end{equation}
where $\overline{n}(z)$ is the average galaxy number density of the 
corresponding unclustered sample at the redshift of each galaxy,
$z_i$ and $z_j$, and 
$J_3(s_{ij}) = \int_0^{s_{ij}} s^2 \xi(s) ds$.
For this integral, we assume a power-law for the correlation function,
$\xi(s) = (s/s_0)^{-\gamma}$,
with parameters $s_0 = 5.59 \hMpc$ and $\gamma = 1.84$, and we integrate
out to the separation $s_{ij}$ of the galaxy pair,
or 30 \hMpc, if the separation is larger than this.
We have checked that the correlation function estimates are insensitive
to the details of the assumed power-law.
If, instead, we assume a power-law consistent with the clustering of
GAMA galaxies in the faint magnitude bin M6,
viz. $s_0 = 3.68 \hMpc$, $\gamma = 1.84$,
individual $w_p(r_\bot)$ estimates change by less than the 1-sigma error bars.
For volume-limited samples, weighting is uniform, i.e. $w_{ij} \equiv 1$.

We then normalise for the relative total numbers of galaxies, $N_g$,
and random points, $N_r$,
by dividing the summed pair weights $DD$, $DR$ and $RR$
for each separation bin by $N_g (N_g - 1)$, $N_g N_r$, and $N_r (N_r - 1)$,
respectively.

\begin{figure}
\includegraphics[width=\linewidth]{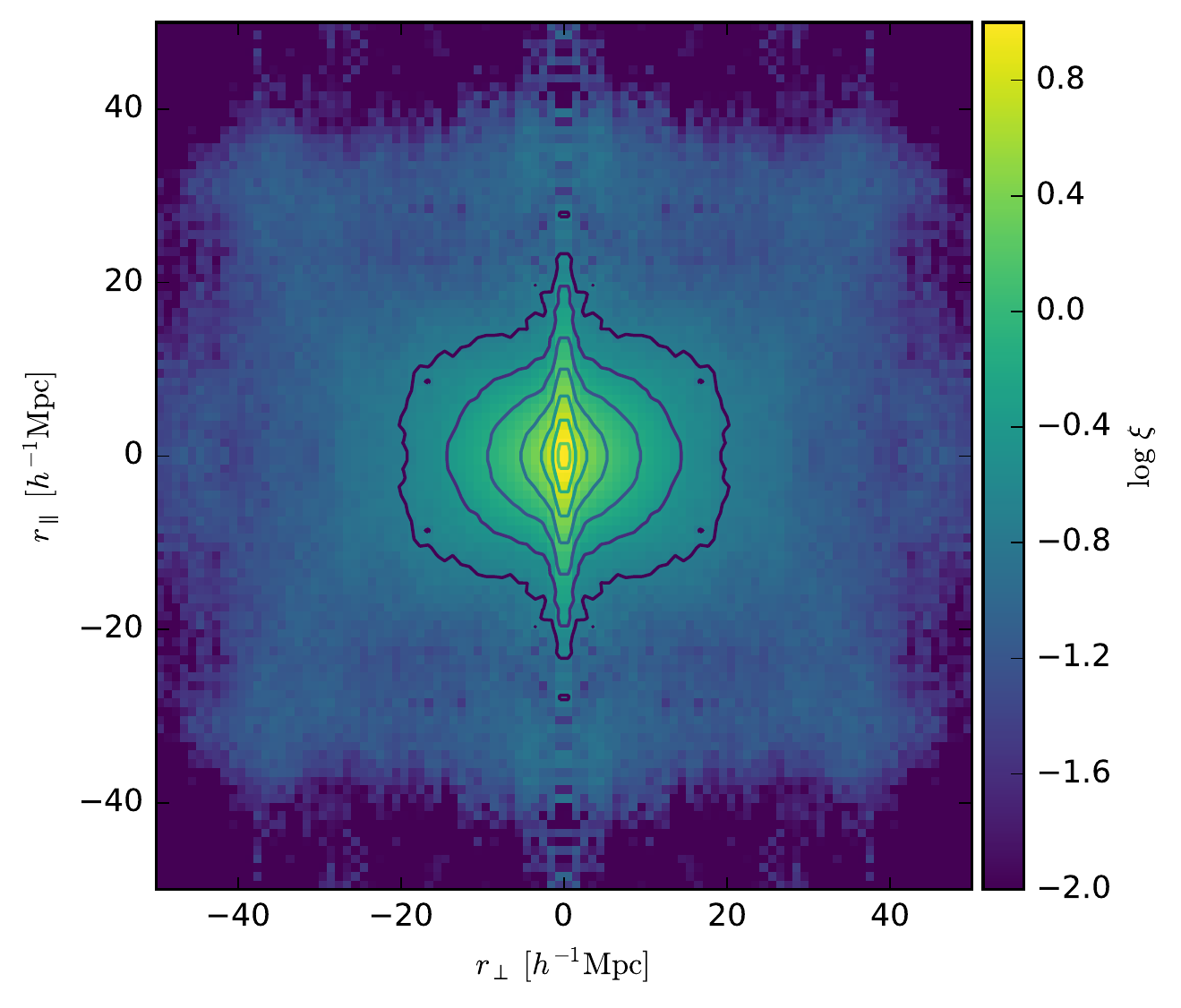}
\includegraphics[width=\linewidth]{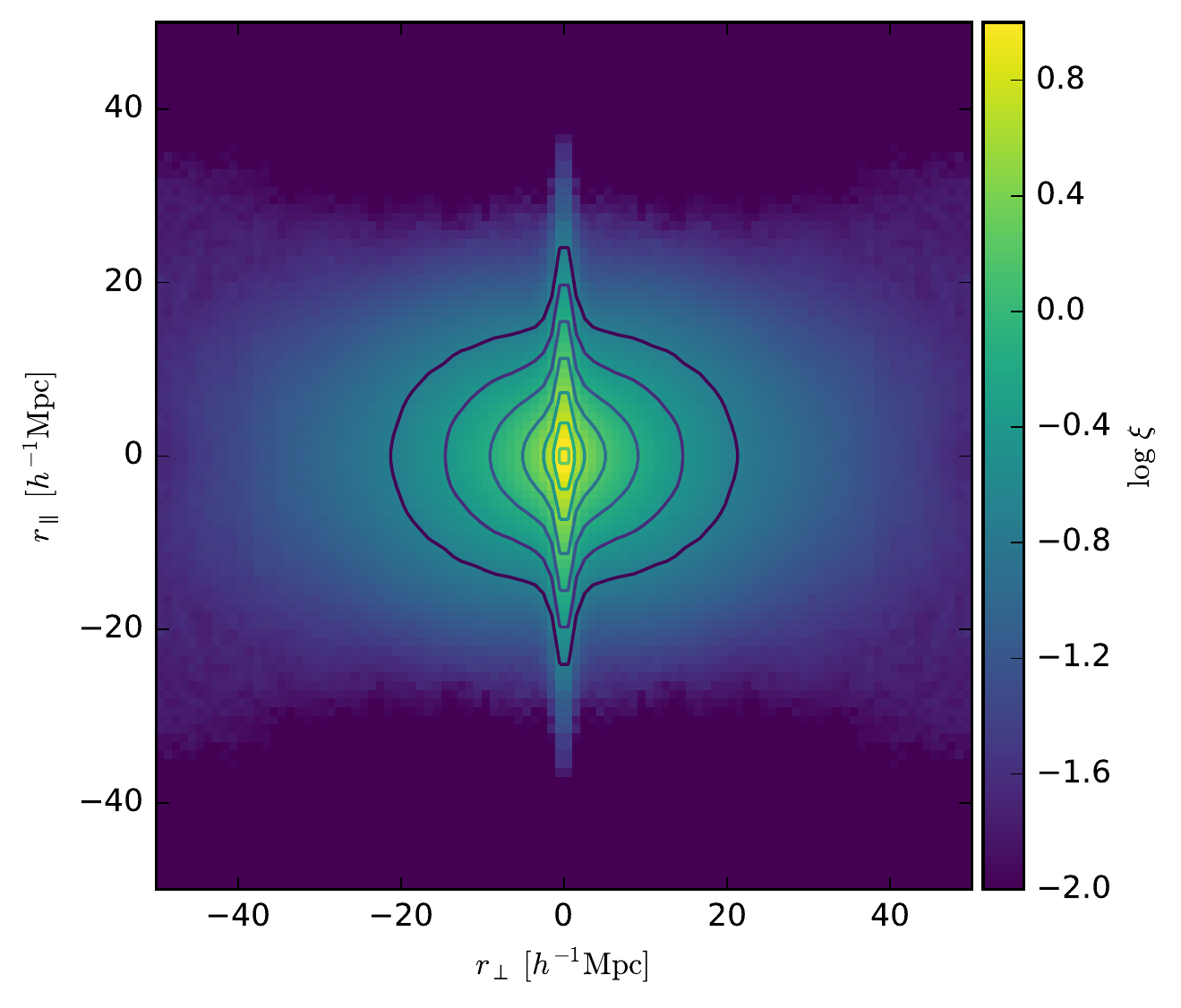}
\caption{The two-dimensional correlation function $\xi(r_\bot, r_\|)$
for $M_r < -20$ volume-limited samples for (top) GAMA-II galaxies and (bottom)
the average from 26 mock catalogues.
To visually clarify departures from isotropy due to peculiar velocities,
and to minimize aliasing effects when the Fourier transform is taken,
the clustering signal is reflected about both axes.
Following \citet{Li2006b}, contour levels increase by factors of 2
from $\xi = 0.1875$ to $\xi = 48$.
}
\label{fig:xi2d}
\end{figure}

The two-dimensional correlation function for our volume-limited sample of 
GAMA galaxies, along with the average correlation function from 26 mock samples,
are shown in Fig.~\ref{fig:xi2d}.
Elongation of the clustering signal along the LOS ($r_\|$-axis) 
at small projected separations, $r_\bot \la 5 \hMpc$, 
and a compression of the LOS
clustering signal at larger separations are both clearly visible.
For this plot, we have calculated $\xi(r_\bot, r_\|)$ in $1 \hMpc$ bins
in both coordinates.
We wish to determine the PVD measurement on the smallest possible scales, 
fully exploiting the high spectroscopic completeness of the GAMA survey.
When estimating the PVD, we thus
measure $\xi(r_\bot, r_\|)$ in logarithmically-spaced bins in both
directions, with $\log_{10} (r/\hMpc)$ ranging from $-2$ to 2 in 20 bins.

Determination of the PVD also requires that
the real-space correlation function $\xi_r(r)$ be known.
We estimate $\xi_r(r)$ from the data via the projected correlation function, 
$w_p(r_\bot)$, which is obtained in the usual way by integrating the observed
two-dimensional correlation function $\xi(r_\bot, r_\|)$
along the LOS direction $r_\|$:
\begin{equation}
  \label{eq:proj}
  w_p(r_\bot) = 2 \int_0^{{r_\|}_{\rm max}} \xi(r_\bot, r_\|) d  r_\|.
\end{equation}
We use an upper integration limit of ${r_\|}_{\rm max} = 40 \hMpc$;
see Appendix~\ref{sec:xi_test} for justification of this choice.

The real-space correlation function $\xi_r(r)$ may then be obtained by
performing the inversion
\begin{equation}
  \label{eq:real}
  \xi_r(r) = -\frac{1}{\pi} \int_r^\infty w_p(r_\bot) (r_\bot^2 - r^2)^{-1/2} d r_\bot.
\end{equation}
This integral is evaluated by linearly interpolating between the binned
$w_p(r_\bot)$ values \citep{Saunders1992}; we use 20 logarithmically spaced
bins from 0.01 to 100 \hMpc.
Since this estimate of $\xi_r(r)$ can be be rather noisy,
we also approximate $\xi_r(r)$ using a power-law fit to the projected 
correlation function $w_p(r_\bot)$ over the separation range 
$0.01 \hMpc\ < r_\bot < 5 \hMpc$.
For a power-law fit, $\xi_r(r) = (r/r_0)^{-\gamma}$, equation~(\ref{eq:real})
yields \citep{Davis1983},
\begin{align}
w_p(r_\bot) &= A r_\bot^{1 - \gamma} \nonumber \\
A &= r_0^\gamma \Gamma(1/2) \Gamma[(\gamma - 1)/2)]/\Gamma(\gamma/2),
\label{eq:power-law}
\end{align}
where $\Gamma$ is the standard gamma function.

\section{Modelling the correlation function} \label{sec:dist}

Historically, two complementary approaches have been taken to model the
two-dimensional galaxy correlation function
in the presence of galaxy peculiar motions,
the `streaming' and `dispersion' models.
In this Section, we briefly review these two models and then proceed
to demonstrate that the peculiar velocity distribution function at small
scales is reasonably well-fit by an exponential function for GAMA galaxies.

\subsection{Streaming model}

In the streaming model
\citep[e.g.][]{Peebles1980,Peebles1993,Davis1983,Fisher1995,Zehavi2002},
$\xi(r_\bot, r_\|)$ is given
by a convolution of the isotropic real-space correlation function $\xi_r(r)$
with the pairwise LOS velocity distribution $f(v_{12})$:
\begin{equation}
\label{eqn:stream}
1 + \xi(r_\bot, r_\|) = H_0 \int_{-\infty}^{\infty} 
\left[ 1 + \xi_r(r) \right] f(v_{12}) dy.
\end{equation}
Here, $y$ is the true LOS separation of the galaxy pair,
the total true separation is $r = \sqrt{r_\bot^2 + y^2}$,
and $v_{12} = H_0(r_\| - y)$ is the relative LOS peculiar velocity.

The pairwise velocities are most often assumed to follow an
exponential distribution:
\begin{equation}
\label{eqn:v_exp}
f_e(v_{12}) = \frac{1}{{\sqrt 2} \sigma_{12}(r_\bot)} 
\exp\left(-\frac{{\sqrt 2}|v_{12} - \vbar|}{\sigma_{12}(r_\bot)}\right),
\end{equation}
or a Gaussian distribution
\begin{equation}
\label{eqn:v_gauss}
f_G(v_{12}) = \frac{1}{\sqrt{2 \pi} \sigma_{12}(r_\bot)} 
\exp\left(-\frac{(v_{12} - \vbar)^2}{2 \sigma^2_{12}(r_\bot)}\right).
\end{equation}

The mean relative peculiar velocity of galaxies separated by a distance $r$ 
(by symmetry directed along the separation vector ${\bm r}$) 
is given by $\overline{\bm v}_{12} = -H_0 g(r) {\bm r}$,
and thus $\vbar = -H_0 g(r) y$ is the LOS component 
of this mean velocity.
As discussed by \citet[p~478]{Peebles1993}, 
one expects $g(r)$ to be close to unity on small scales where the 
peculiar velocity cancels out the Hubble flow within bound structures.
At larger scales, $g(r)$ should tend to zero as uncorrelated galaxies
move with the Hubble flow.
In this work, 
we use the expression given by 
\citet[equation~6, hereafter `JSD model']{Juszkiewicz1999}
for the mean radial pairwise velocity $\vbar(r)$.

The streaming model has been developed and improved by a number of authors,
including
\citet{Fisher1995,Sheth1996,Scoccimarro2004,Reid2011,Bianchi2015,Uhlemann2015},
in order to find a more complete description of the velocity distribution
function $f(v_{12})$.
The main focus of these works has been to improve the model in the linear
regime.
In the present work, we are focused on strongly non-linear scales,
for which we show that assuming an exponential velocity distribution
provides a good fit to simulations,
provided that $\pvd(k)$ is allowed to vary with scale.

\subsection{Dispersion model}

Rather than assuming a model for the mean streaming velocity
$\vbar(r)$, the dispersion model combines the \citet{Kaiser1987} 
linear infall model with an assumed small-scale velocity distribution function.
In Fourier space, the redshift space power spectrum $P_s(k, \mu)$ may be related
to the real space power spectrum $P_r(k)$ by 
\citep{Peacock1994,Cole1995,Jing2001a}\footnote{See \citet{Scoccimarro2004}
  for an improved version of the dispersion model that allows for coupling
  between the velocity and density fields.}
\begin{equation} \label{eqn:p_mod}
P_s(k, \mu) = P_r(k) (1 + \beta \mu^2)^2 D(k \mu \pvd(k)),
\end{equation}
where $\mu$ is the cosine of the angle between the wavevector $\bm k$ 
and the LOS.
The factor $(1 + \beta \mu^2)^2$ is the \citet{Kaiser1987} linear compression
effect\footnote{The linear redshift distortion parameter 
$\beta = f(\Omega_m)/b$,
where $f(\Omega_m) \approx \Omega_m^{0.6}$ is the dimensionless growth rate
of structure in the linear regime and $b$ is the galaxy bias parameter.}
and the factor $D$ is the damping caused by random motions of galaxies.
For an exponential form of the pairwise velocity distribution 
(equation~\ref{eqn:v_exp}), its Fourier transform $D$ is a Lorentzian
\begin{equation}
D(k \mu \pvd(k)) = [1 + 0.5(k \mu \pvd(k))^2]^{-1}.
\end{equation}

The \citet{Kaiser1987} linear infall model has been translated from
Fourier to configuration space by \citet{Hamilton1992}
to predict the shape of the two-dimensional correlation function
$\xi'(r_\bot, r_\|)$ due to coherent infall:
\begin{equation}
  \xi'(r_\bot, r_\|) = \xi_0(s) P_0(\mu) + \xi_2(s) P_2(\mu) +
  \xi_4(s) P_4(\mu),
\end{equation}
where the $P_l(\mu)$ are Legendre polynomials and the harmonics of the
correlation function are defined in Appendix~\ref{sec:hamilton}.
One can then approximate the observed two-dimensional correlation function
by convolving $\xi'(r_\bot, r_\|)$ with the peculiar velocity distribution, 
e.g. \citet{Hawkins2003},
\begin{equation}
\label{eqn:beta}
\xi(r_\bot, r_\|) = 
\int_{-\infty}^{\infty} \xi'(r_\bot, r_\| - v_{12}/H_0) f(v) dv,
\end{equation}
where $f(v_{12})$ is now assumed to be distributed around zero,
and is thus given by equation~(\ref{eqn:v_exp}) or (\ref{eqn:v_gauss})
with $\overline{v}_{12} \equiv 0$.

\subsection{Pairwise velocity distribution function}

Previous work, e.g. \citet{Loveday1996,Landy1998,Hawkins2003},
has found that the galaxy pairwise velocity distribution is better
fit by an exponential than a Gaussian function.
\citet{Sheth1996} shows that an exponential distribution is expected on
highly non-linear scales from Press--Schechter theory.
We determine the pairwise velocity distribution for the GAMA-II data
using the method of \citet{Landy1998},
also used by \citet{Landy2002} and \citet{Hawkins2003},
which deconvolves the real-space correlation function from the 
peculiar velocity distribution.

\begin{figure}
\includegraphics[width=\linewidth]{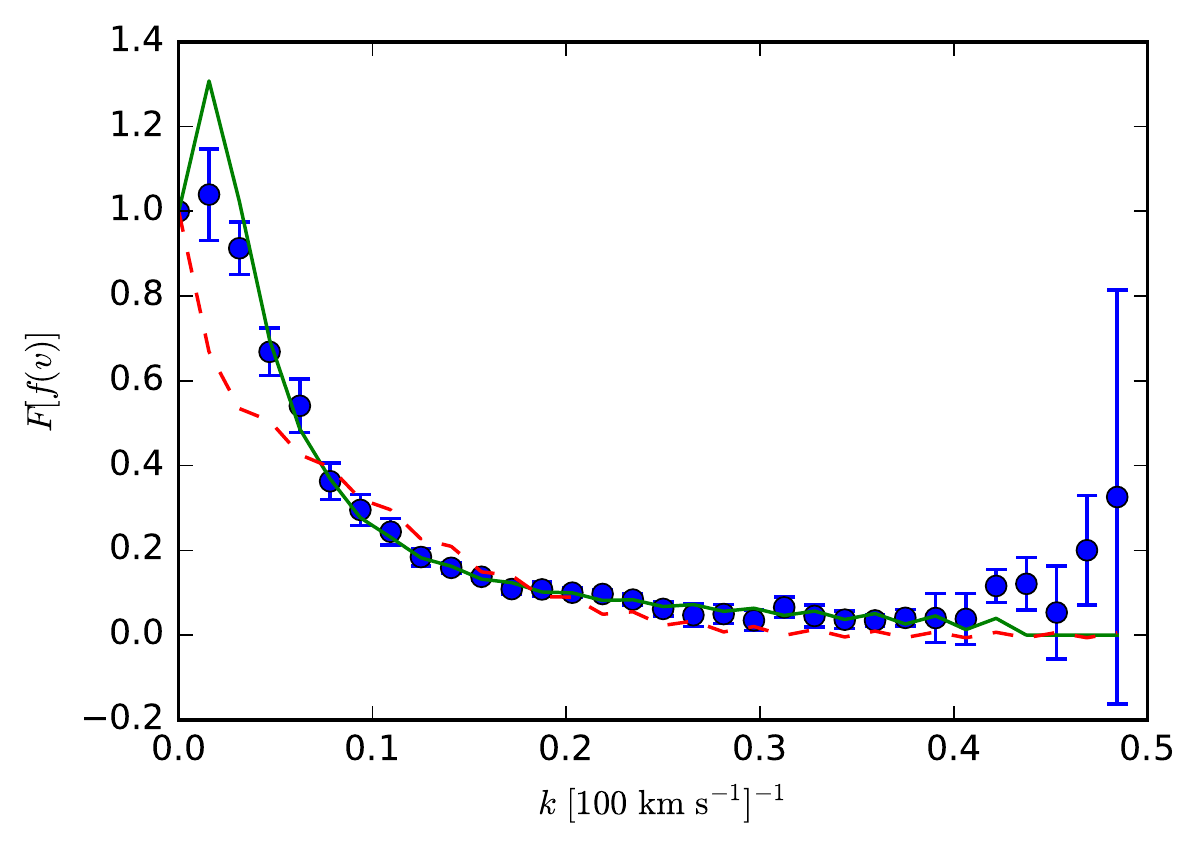}
\includegraphics[width=\linewidth]{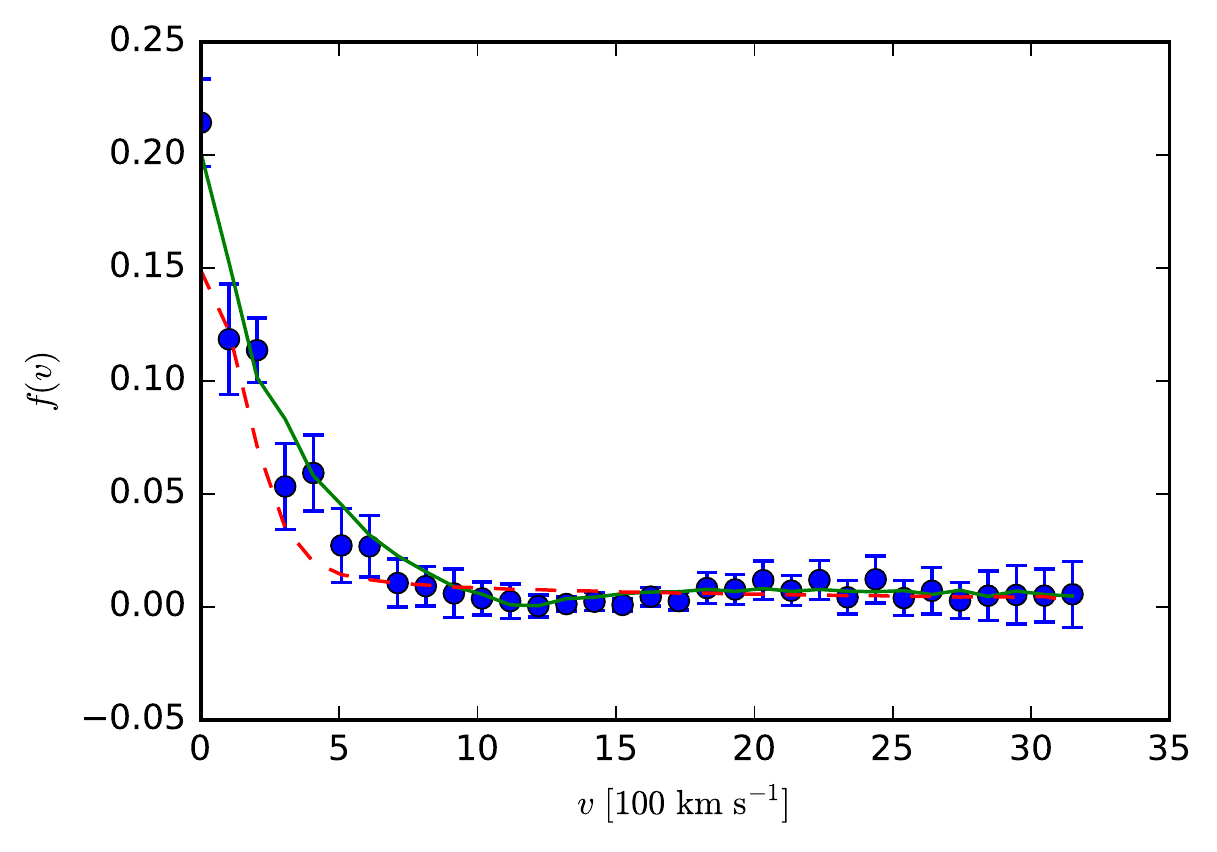}
\caption{The peculiar velocity distribution function (bottom) and its
Fourier transform (top) for the GAMA V0 sample (symbols with errorbars).
The continuous green and red dashed lines show the best fit dispersion model 
predictions with exponential and Gaussian velocity distribution functions 
respectively.
}
\label{fig:vdist}
\end{figure}

\begin{figure}
\includegraphics[width=\linewidth]{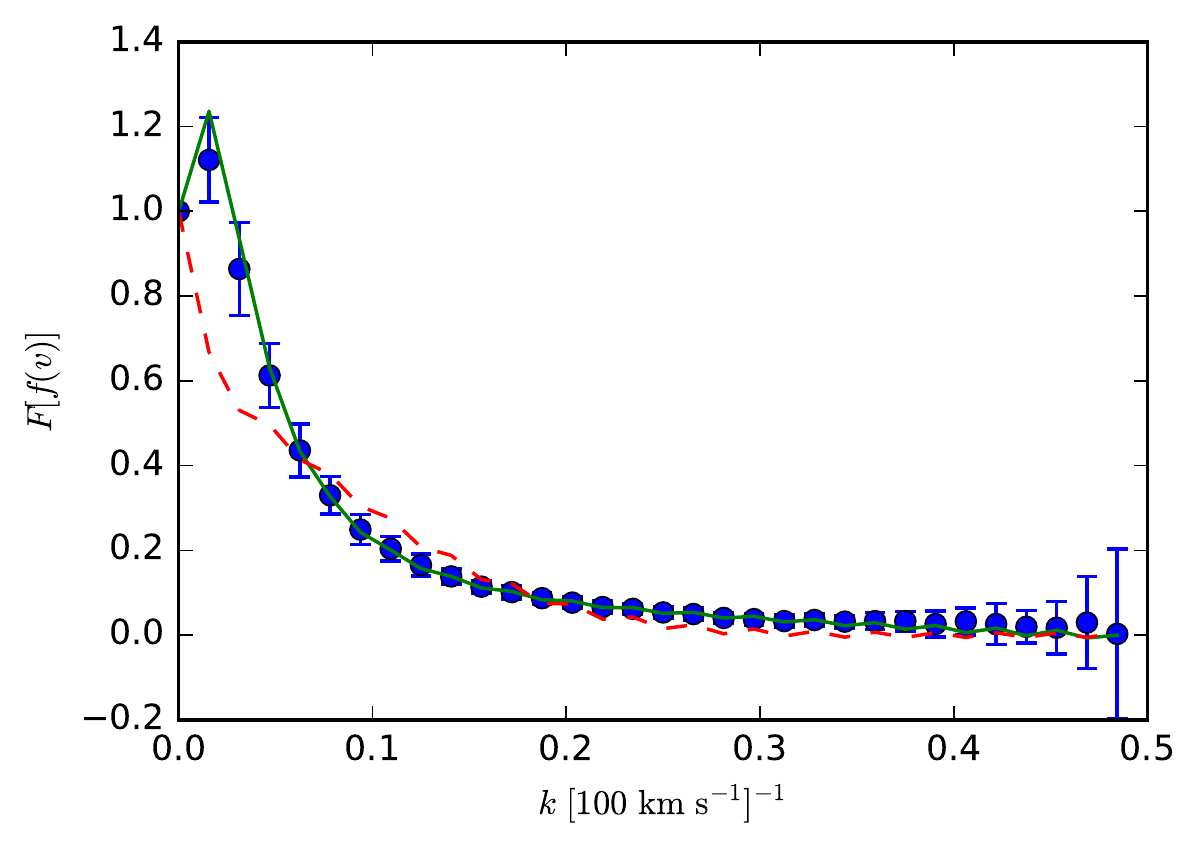}
\includegraphics[width=\linewidth]{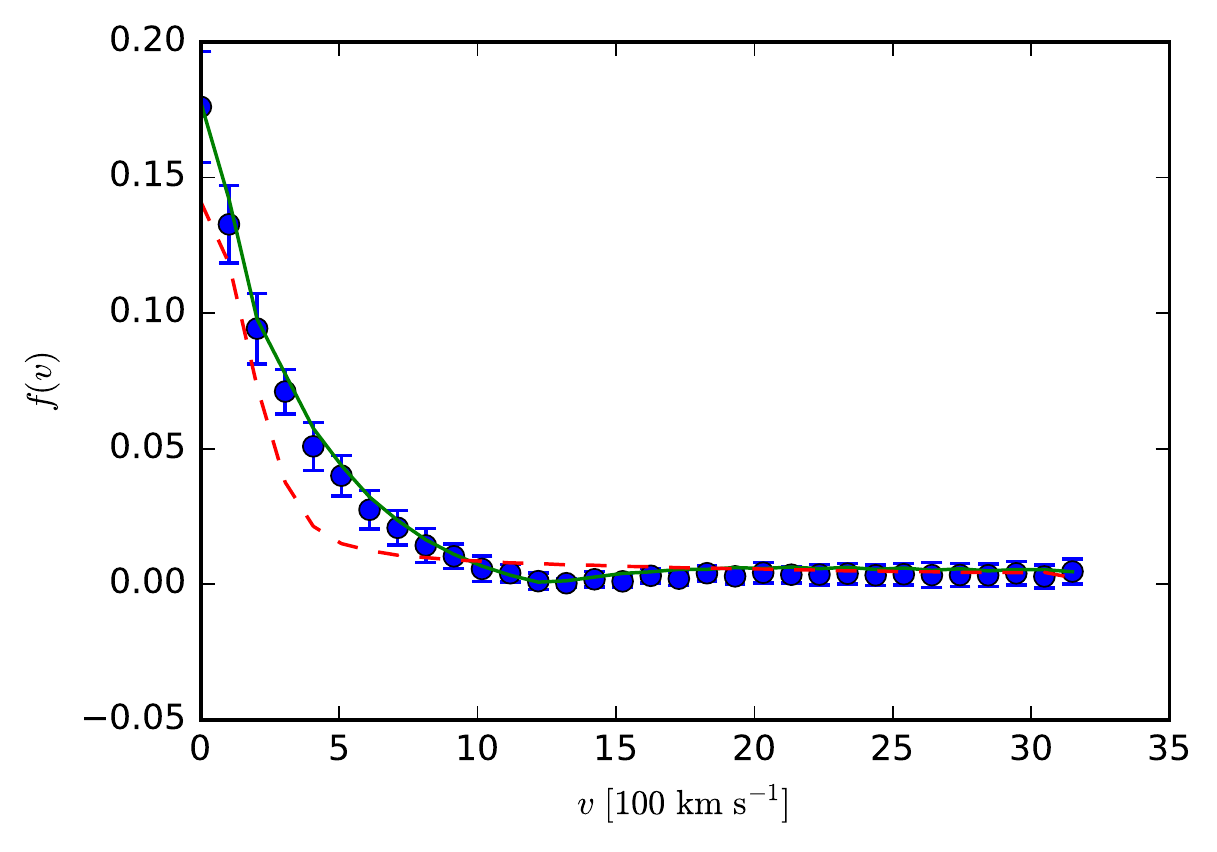}
\caption{As Fig.~\ref{fig:vdist} but from the average of 26 mock catalogues.
}
\label{fig:mock_vdist}
\end{figure}

\begin{table*}
 \caption{Fits of $\beta$ infall parameter and velocity dispersion $\sigma_{12}$
(in \kms) to the Fourier transformed velocity distribution function for both 
the GAMA V0 sample and the mean of the 26 mock V0 samples.
For all fits there are 29 degrees of freedom.
}
 \label{tab:vdist}
 \begin{math}
 \begin{array}{lrrrrrr}
 \hline
& \multicolumn{3}{c}{\mbox{GAMA}} & \multicolumn{3}{c}{\mbox{Mocks}} \\
\multicolumn{1}{c}{\mbox{Model}} & \multicolumn{1}{c}{\beta} & 
\multicolumn{1}{c}{\sigma_{12}} & \multicolumn{1}{c}{\chi^2} &
\multicolumn{1}{c}{\beta} & 
\multicolumn{1}{c}{\sigma_{12}} & \multicolumn{1}{c}{\chi^2} \\
\hline
\mbox{Exponential} & 0.75 \pm 0.03 & 676 \pm 34 & 32 & 0.68 \pm 0.05 & 690 \pm 71 & 5 \\
\mbox{Gaussian} & -0.88 \pm 0.08 & 106 \pm 12 & 178 & -0.92 \pm 0.04 & 111 \pm 10 & 112 \\
 \hline
 \end{array}
 \end{math}
\end{table*}

We take the 2D Fourier transform\footnote{We utilise the Numerical Python
  discrete Fourier transform package {\tt numpy.fft}.}
of the $\xi(r_\bot, r_\|)$ grid measured out to 32 \hMpc\ 
to give $\hat{\xi}(k_\bot, k_\|)$.\footnote{The true power spectrum
  will only be obtained by taking the Fourier transform
  of the correlation function measured on all scales.
  We have verified that we get consistent results in this section when
  starting with $\xi(r_\bot, r_\|)$ measured out to 64 \hMpc.}
By the slicing--projection theorem \citep{Landy1998}, cuts of 
$\hat{\xi}(k_\bot, k_\|)$ along the $k_\bot$ and $k_\|$ axes, i.e.
$\hat{\xi}(k_\bot, 0)$ and $\hat{\xi}(0, k_\|)$, are equivalent to the
Fourier transforms of the real-space projections of $\xi(r_\bot, r_\|)$
onto the $r_\bot$ and $r_\|$ axes.
The projection onto the $r_\bot$ axis is distortion-free, whereas
the projection onto the $r_\|$ axis gives the real-space correlation function
convolved with the peculiar velocity distribution function.
The ratio ${\cal F}[f(v_{12})](k) = \hat{\xi}(k_\bot=k, 0)/\hat{\xi}(0, k_\|=k)$ 
is thus the Fourier transform of the peculiar velocity distribution function;
taking the inverse transform of this ratio yields $f(v_{12})$.

We fit the dispersion model to both the observations and mocks, 
with free parameters $\beta$ and $\sigma$, 
by performing a $\chi^2$ least--squares fit
of the predicted to the measured ${\cal F}[f(v_{12})]$.
We fit to ${\cal F}[f(v_{12})]$ rather than $f(v_{12})$ 
since the covariance matrix of
$f(v_{12})$ estimates shows significant anti-correlations between 
odd and even-numbered bins.
Both ${\cal F}[f(v_{12})]$ and the inferred $f(v_{12})$ for GAMA and mock
volume-limited samples
are shown in Figs.~\ref{fig:vdist} and \ref{fig:mock_vdist}, 
along with best-fit predictions from
dispersion models with exponential and Gaussian velocity distribution functions.
The best-fit parameters and $\chi^2$ values are given in Table~\ref{tab:vdist}.
It is clear that the exponential distribution function provides a much better
fit to both the GAMA data and the mocks than the Gaussian distribution,
which also gives unphysical, negative values for $\beta$.
We therefore assume an exponential form for $f(v_{12})$ for the rest 
of this paper.

While this method is useful for determining the shape of the distribution 
function $f(v_{12})$, it is not ideal for estimating the values of $\beta$
and the velocity dispersion $\sigma_{12}$, since it averages over a range 
of linear and non-linear separations.
In fact, the value of $\beta$ obtained for the mock V0
samples is completely inconsistent with the expected value of
$\beta \approx 0.37$ (Sec.~\ref{sec:subsamples}).
The uncertainties quoted in Table~\ref{tab:vdist} account only for
scatter between realisations,
and not for systematic errors due to inadequacies in 
the dispersion model when applied over a wide range of scales.
In the following section we obtain separation-dependent estimates of the 
velocity dispersion $\sigma_{12}(r_\bot)$ by fitting directly to the 
$\xi(r_\bot, r_\|)$ grid.

\section{Mock tests of PVD estimators} \label{sec:method}

In this section, we test three methods for recovering the PVD 
from observational data,
two based on the correlation function in redshift space, via the streaming
and dispersion models, and one based on the dispersion model in Fourier space.
All start with the 2d correlation function $\xi(r_\bot, r_\|)$,
measured as described in Section~\ref{sec:xi2d},
and are based on least-squares fitting of a model 2d correlation function
or power spectrum to the measured one.
The $\xi(r_\bot, r_\|)$ values are strongly correlated, and so one
should in principle use the full covariance matrix or its principal components
\citep[e.g.][]{Norberg2009} in least-squares fitting.
In practice, even when using mock catalogues, 
for which a covariance matrix may be reasonably well-determined,
improvements over using just the diagonal elements are negligible, at best.
With only nine jackknife samples for GAMA data, covariance matrix estimates 
are even more susceptible to noise.
Therefore all fitting to data is done using only
diagonal elements of the covariance matrix.
We show in Section~\ref{sec:beta-test} below that this introduces only a small
bias in the inferred PVD using the dispersion model.

Before describing these clustering-based estimators of the PVD, 
we first discuss a direct
measurement of the PVD from the mock catalogues, which will be used to
test the veracity of the clustering-based estimates.

\subsection{Direct mock-PVD measurement}

Using the observed ($z_{\rm obs}$) and cosmological ($z_{\rm cos}$) redshifts
provided in the mock galaxy catalogues, 
one can determine the LOS peculiar velocity ($v_{\rm pec}$) 
for each galaxy using \citep{Harrison1974}
\begin{equation}
1 + v_{\rm pec}/c = (1 + z_{\rm obs})/(1 + z_{\rm cos}).
\end{equation}
We then define the relative LOS velocity for a pair of galaxies
as $v_{12} = v_{\rm pec, 2} - v_{\rm pec, 1}$,
where galaxy 1 is the closer of the pair, so that $v_{12}$ is negative
for galaxies that are approaching each other.
Note that this formula is only accurate for galaxy pairs at small 
angular separation, and so we limit our measurements to pairs of galaxies
separated by less than $12^\circ$ (the RA extent of each GAMA region).

\begin{figure}
\includegraphics[width=\linewidth]{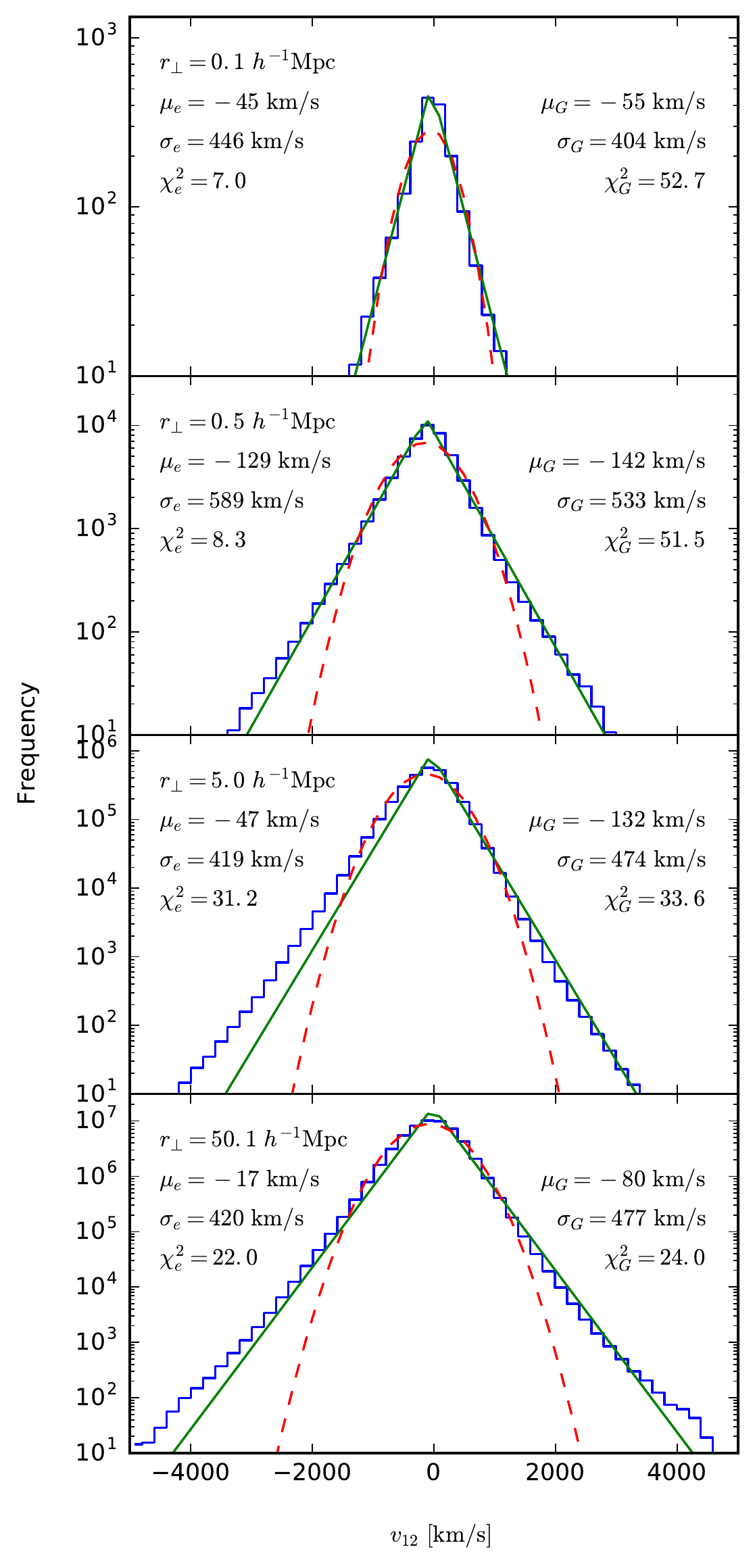}
\caption{LOS relative velocity distributions for mock galaxy pairs
(stepped histograms), exponential fits (continuous lines) and
Gaussian fits (dashed lines) in four bins of projected separation as labelled.
We include all pairs along the line of sight direction up to $r_\| = 50 \hMpc$.
Fit parameters and reduced $\chi^2$ values are shown for the exponential fits
on the left and for Gaussian fits on the right.
}
\label{fig:pvd_mock_hist}
\end{figure}

The LOS pairwise velocity distributions for mock galaxy pairs 
in four representative bins of projected separation,
including all pairs along the line of sight direction up to $r_\| = 50 \hMpc$,
are shown in Fig.~\ref{fig:pvd_mock_hist}.
Visually, the distributions are reasonably well-fit by 
exponential functions for galaxy pairs at projected separation 
$r_\bot \la 1 \hMpc$ (e.g. top two panels), even though the reduced $\chi^2$ 
values formally rule out an exponential fit (the standard error from the
scatter between mock realisations is tiny).
At larger separations, a growing skewness in the distributions
toward negative velocities\footnote{This skewness in the
pairwise velocity distribution has previously been reported from simulations
by \citet{Juszkiewicz1998} and \citet{Magira2000}.
See \citet{Bianchi2015,Bianchi2016} for a bivariate Gaussian description
for the pairwise velocity distribution function that can account for this 
asymmetry.}, 
and decreasing random errors, result in poorer fits.
It is clear that a Gaussian grossly under-fits the tails of the distribution
at all separations.

For pairs of galaxies in logarithmically-spaced separation bins 
$(r_\bot, r_\|)$ in redshift-space, we calculate the
mean and standard deviation of the pairwise LOS velocity distribution,
$\vbar$ and $\sigma_{12}$ respectively, as well as the 
maximum-likelihood velocity dispersion $\sigma^{\rm exp}_{12}$ for an 
exponential distribution (equation~\ref{eqn:v_exp}), namely
\begin{equation}
\sigma^{\rm exp}_{12} = \frac{\sqrt{2}}{N} \sum_{i=1}^N |v_{12,i} - \vbar|,
\end{equation}
where the sum is carried out over all $N$ galaxy pairs in each separation bin.

\begin{figure}
\includegraphics[width=\linewidth]{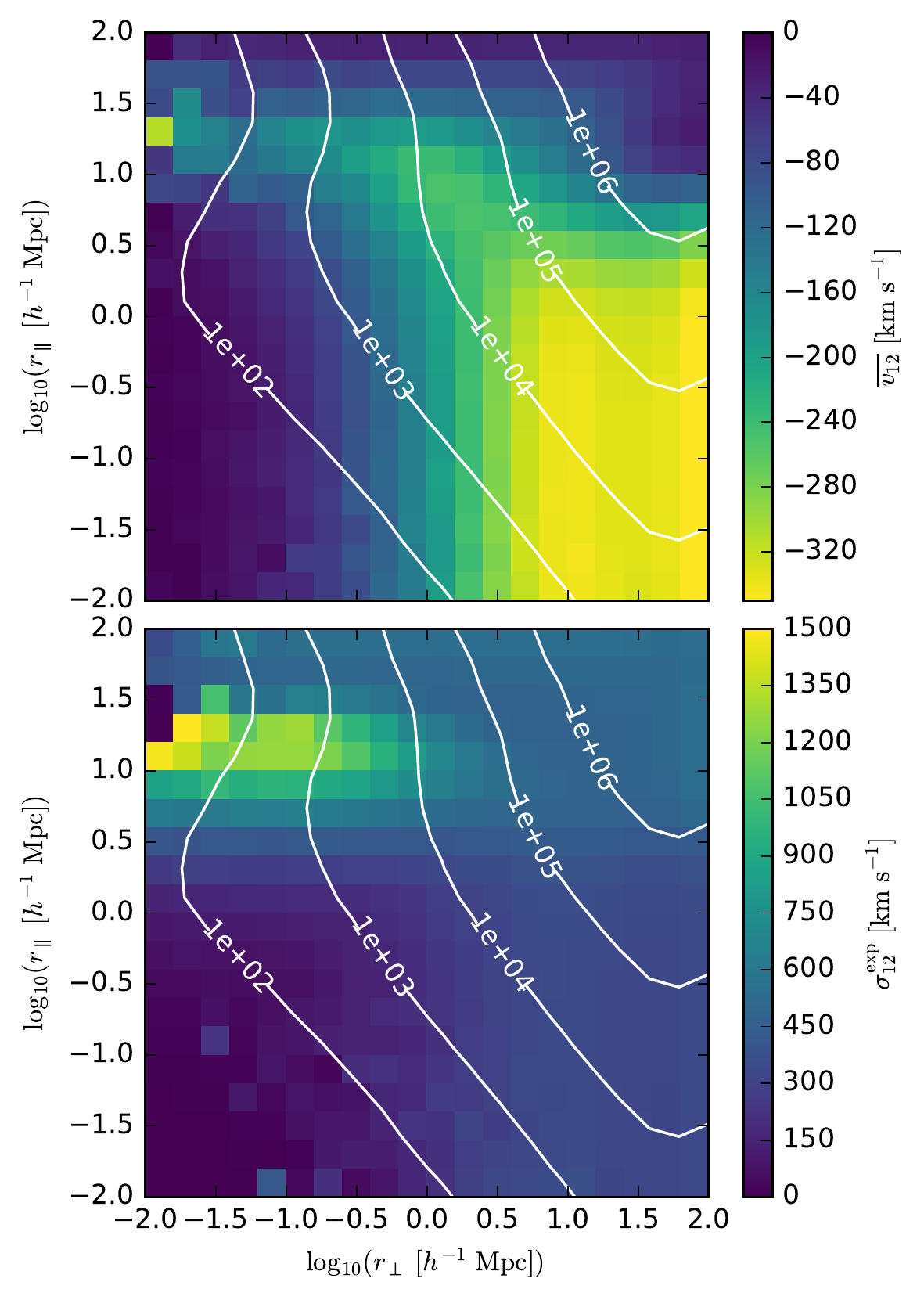}
\caption{Mean (top) and exponential dispersion (bottom) of the mock LOS 
pairwise velocities, in logarithmic bins of redshift space separation.
Contour lines from left to right connect bins containing 
$10^2, 10^3, 10^4, 10^5$ and $10^6$ galaxy pairs.
}
\label{fig:pvd_mock_twod}
\end{figure}

In Fig.~\ref{fig:pvd_mock_twod}, we show the mean and exponential
dispersion of the mock LOS pairwise velocities in bins of two-dimensional
redshift-space separation.
It is interesting to see that the largest (negative) mean velocities
and dispersions occur at very different separation bins.
The most negative mean velocities of $\vbar \la -300 \kms$
are seen at projected separation $r_\bot \ga 3 \hMpc$ 
and LOS separation $r_\| \la 3 \hMpc$.
The highest velocity dispersions, 
of up to $\sigma^{\rm exp}_{12} \approx 1500 \kms$,
are seen at $r_\bot \la 1 \hMpc$ and $10 \la r_\| \la 30 \hMpc$.
The PVD is uniformly low, $\sigma^{\rm exp}_{12} \la 300 \kms$,
at LOS separations 
$r_\| \la 5 \hMpc$ regardless of projected separation $r_\bot$.

Values of $\vbar(r_\bot)$ and $\sigma^{\rm exp}_{12}(r_\bot)$
as a function of projected separation $r_\bot$ alone are obtained by
averaging over $r_\|$ separation bins, weighting by the number of
galaxy pairs per bin.
It is clear that most of the contribution to the velocity dispersion
$\sigma^{\rm exp}_{12}(r_\bot)$
will come from LOS separations $5 \la r_\| \la 30 \hMpc$.
The estimated dispersion is insensitive to the upper limit of 
LOS separation beyond $r_\| \approx 30 \hMpc$ due to the modest
drop in $\sigma^{\rm exp}_{12}$ values at larger $r_\|$.
The estimated mean velocity $\vbar(r_\bot)$ {\em is} 
sensitive to the limiting
$r_\|$ due to the steady decline in $|\vbar|$ beyond 
$r_\| \ga 30 \hMpc$, so that the amplitude of $\vbar(r_\bot)$
decreases with increasing ${r_\|}_{\rm max}$.
Since galaxies with LOS separation $r_\| \ga 40 \hMpc$
($\Delta v \ga 4000 \kms$) are unlikely to be correlated,
we impose an upper limit of  $r_\| = 40 \hMpc$ when calculating
$\vbar(r_\bot)$ and $\sigma^{\rm exp}_{12}(r_\bot)$.

Note that it is unlikely that the mock catalogues have sufficient resolution
to reliably predict galaxy dynamics on scales below about $0.1 \hMpc$.
Nevertheless, by comparing clustering-inferred estimates of the PVD
with those obtained directly from the peculiar velocity information,
we can still use the mocks to test our methods down to scales
$r_\bot = 0.01 \hMpc$.

\begin{figure}
\includegraphics[width=\linewidth]{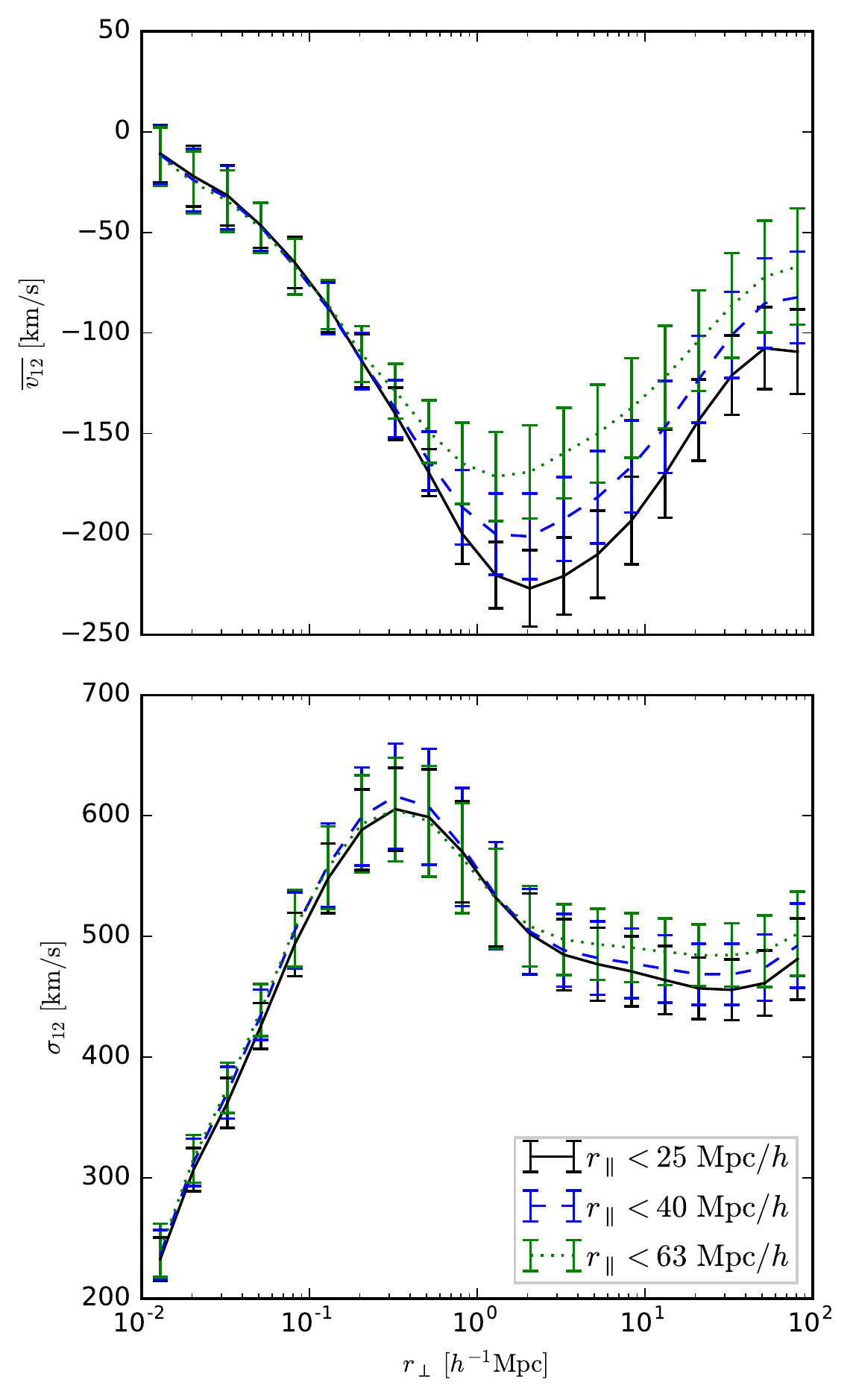}
\caption{LOS mean pairwise velocity $\vbar$ (upper panel),
and  velocity dispersion $\sigma_{12}$ (lower panel)
for pairs of mock galaxies as a function of projected separation.
The different line styles and colours represent different upper limits
to projected separation $r_\|$ as indicated in the lower panel.
}
\label{fig:pvd_mock}
\end{figure}

Maximum-likelihood estimates of mean pairwise velocity and exponential
velocity dispersion are shown in Fig.~\ref{fig:pvd_mock} 
as a function of projected separation $r_\bot$,
including pairs with LOS separation up to three different values of 
$r_\|$ as indicated in the lower panel.
We see net LOS infall between pairs of galaxies on all scales, 
with maximum infall of $\vbar(r_\bot) \approx -200 \kms$ 
at separation $r_\bot \approx 2 \hMpc$ with ${r_\|} < 40 \hMpc$.
Exponential velocity dispersion rises from 
$\sigma^{\rm exp}_{12}(r=0.01 \hMpc) \approx 200 \kms$, peaking at 
$\sigma^{\rm exp}_{12}(r \approx 0.4 \hMpc) \approx 600 \kms$
and tending to $\sigma^{\rm exp}_{12}(r \approx 100 \hMpc) \approx 500 \kms$ 
on large scales.
Note that the velocity dispersion estimates are insensitive to the upper limit
of projected separation.

Having made a direct measurement of the PVD from the mock catalogues,
we can now investigate PVD estimates based on the anisotropy of redshift-space
clustering.

\subsection{Streaming model} \label{sec:stream-test}

\begin{figure}
\includegraphics[width=\linewidth]{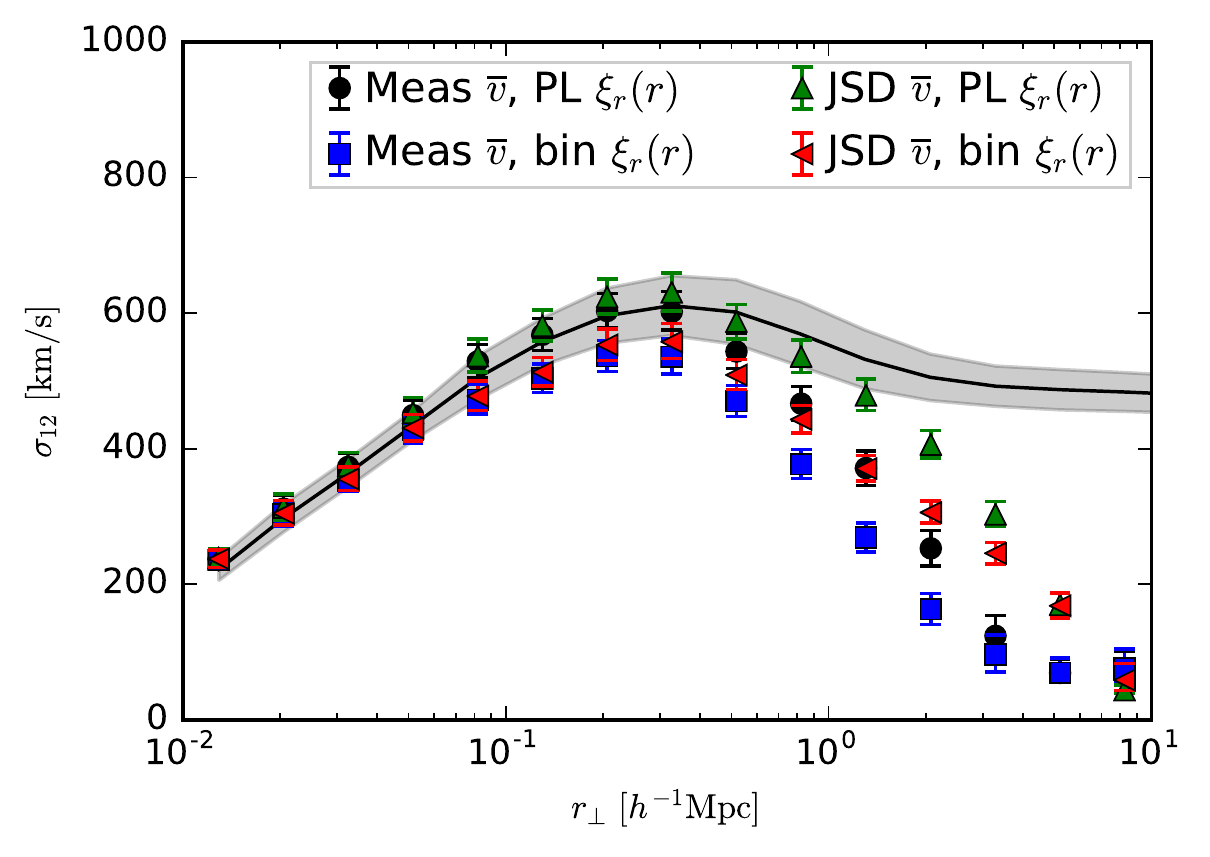}
\caption{Pairwise LOS velocity dispersion estimates from the mock galaxy V0 samples.
The continuous line and shaded region show the direct estimate of 
$\sigma_{12}(r_\bot)$ and its standard deviation reproduced from 
Fig.~\ref{fig:pvd_mock}.
Symbols show PVD estimates recovered by fitting the streaming model
(Section~\ref{sec:stream-test}) to the two-dimensional correlation function.
Black circles and blue squares show results using the measured 
$\vbar(r)$ and power-law and binned measurements of $\xi_r(r)$,
respectively.
Green and red triangles show results using the JSD model 
$\vbar(r)$ and power-law and binned measurements of $\xi_r(r)$,
respectively.
}
\label{fig:mock_stream}
\end{figure}

We find the velocity dispersion $\pvd(r_\bot)$ in bins of 
projected separation $r_\bot$
by least-squares fitting of the observed two-dimensional correlation function 
$\xi(r_\bot, r_\|)$ with the prediction from equation (\ref{eqn:stream}), 
using LOS bins $r_\| < 40 \hMpc$.
We test this estimator for the PVD using the volume-limited mock catalogues 
in Fig.~\ref{fig:mock_stream}.
We compare results obtained using power-law and binned estimates of $\xi_r(r)$,
as well as both directly-measured and the JSD model for mean-streaming
velocities $\vbar(r)$.
All estimates are consistent with the directly-determined PVD on small scales,
$r_\bot \la 0.3 \hMpc$; all tend to underestimate the PVD on larger scales.
Of the four variants of this estimator, that using the JSD model for \vbar\, 
along with a power-law fit for $\xi_r(r)$, performs better than the others,
providing reliable estimates of the PVD to $r_\bot \la 1 \hMpc$.

\subsection{Dispersion model in configuration space} \label{sec:beta-test}

\begin{figure}
\includegraphics[width=\linewidth]{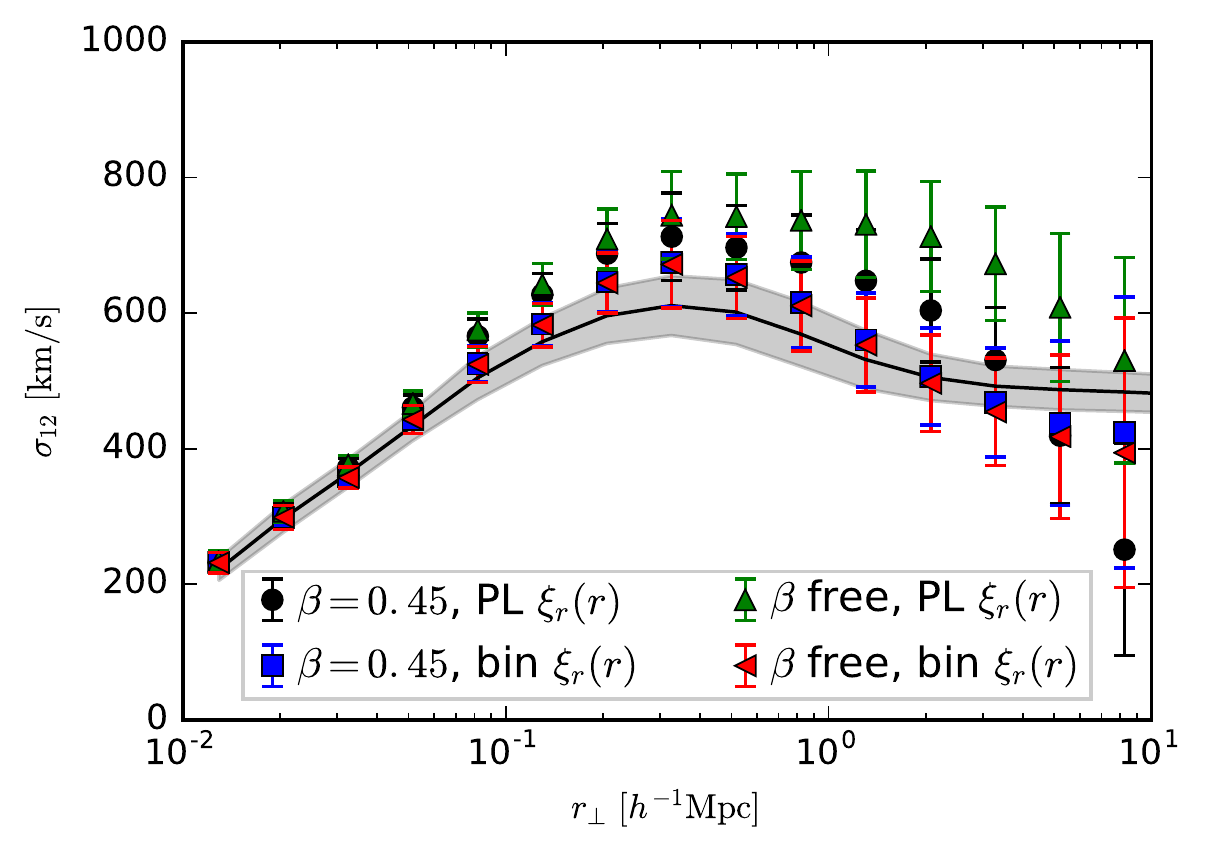}
\caption{As Fig.~\ref{fig:mock_stream} but showing PVD estimates
recovered by fitting the dispersion model
(Section~\ref{sec:beta-test}) to the two-dimensional correlation function.
Black circles show the results of using a power-law fit to $w_p(r_\bot)$
to predict $\xi_r(r)$, blue squares use a numerical inversion of the
binned $w_p(r_\bot)$ measurements to predict $\xi_r(r)$.
Both of these measurements assume a fixed value of $\beta = 0.45$.
Allowing $\beta$ to vary as a free parameter, the power-law $\xi_r(r)$
favours $\beta = 0.66$ (green triangles); the binned $\xi_r(r)$
favours $\beta = 0.43$ (red triangles).
}
\label{fig:mock_beta}
\end{figure}

\begin{figure}
\includegraphics[width=\linewidth]{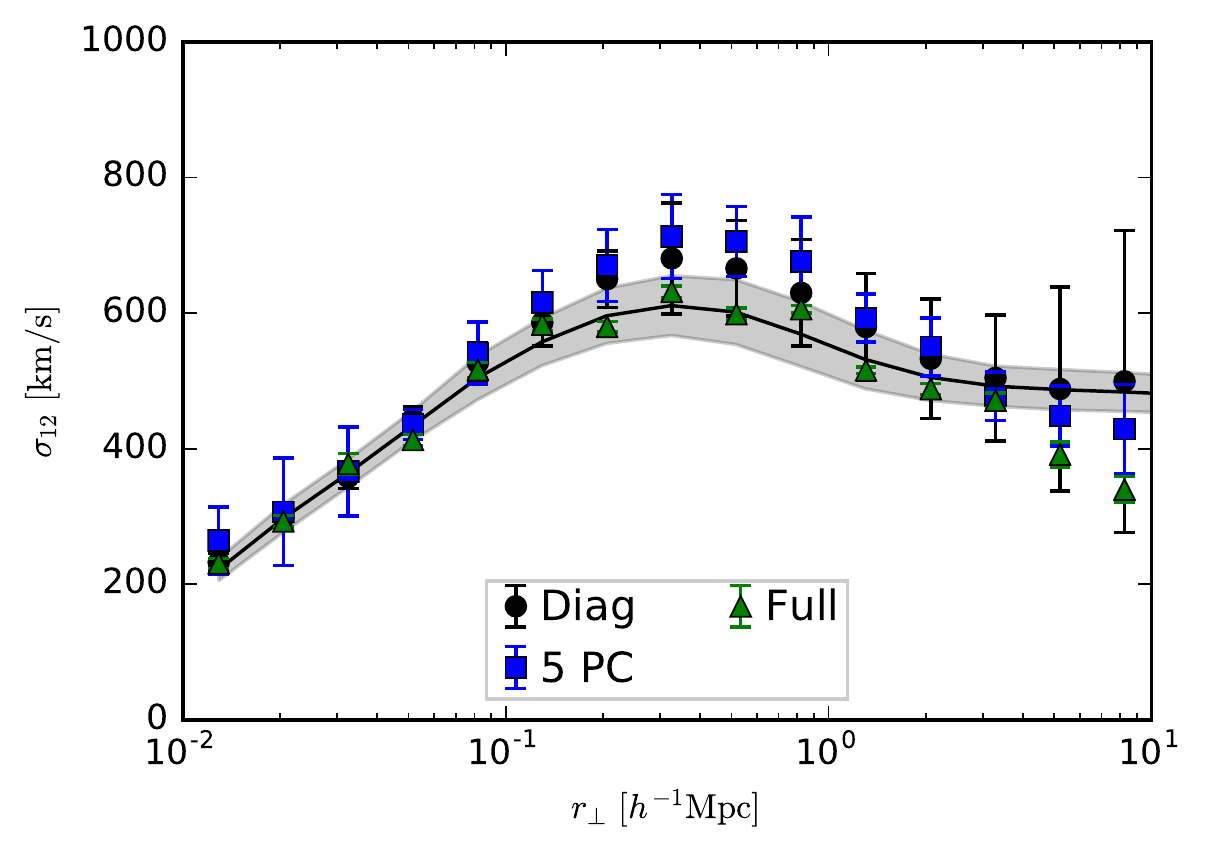}
\caption{Dispersion model estimates of the PVD using the full covariance matrix of
  $\xi(r_\bot, r_\|)$ (green triangles),
  the diagonal components only (black circles),
  or the first five principal components (blue squares).
  We use binned $\xi_r(r)$ measurements and hold $\beta$ fixed at $\beta = 0.5$.
}
\label{fig:mock_beta_cov}
\end{figure}

We find the velocity dispersion $\pvd(r_\bot)$ in bins of 
projected separation $r_\bot$
by least-squares fitting of the observed two-dimensional correlation function 
$\xi(r_\bot, r_\|)$ with the prediction from equation (\ref{eqn:beta}), 
using LOS bins $r_\| < 40 \hMpc$.
We test this estimator for the PVD using the volume-limited mock catalogues 
in Fig.~\ref{fig:mock_beta}.
This plot shows results using both power-law fits and
binned measurements of the real-space correlation function $\xi_r(r)$,
and also with $\beta$ fixed at $\beta = 0.45$ or allowed to vary as a 
free parameter.
It is clear that the binned estimates of $\xi_r(r)$ 
(blue squares and red triangles for fixed and free $\beta$ respectively) 
give a more reliable measure
of the PVD over a wider range of scales than a single power-law fit to
$w_p(r_\bot)$ (black circles and green triangles).
The former measurements lies within $\approx 1 \sigma$ of the 
directly-determined PVD on all scales from 0.01 to 10 $\hMpc$.

Fig.~\ref{fig:mock_beta_cov} explores the effects of using the full
covariance matrix of the correlation function measurements
$\xi(r_\bot, r_\|)$, just the diagonal components,
or the first five principal components, when fitting the dispersion model.
While we see that using the full covariance matrix improves the estimates
on scales $r_\| \la 1 \hMpc$, they are worsened on larger scales.
The full covariance matrix estimates for the GAMA data will be nosier
than those for the 26 mock realisations, and so for our main results
we use only the diagonal elements.

\subsection{Dispersion model in Fourier space}
\label{sec:pvdf}

We determine the redshift space power spectrum $P_s(k, \mu)$ 
from the two-dimensional correlation function $\xi(r_\bot, r_\|)$ using
the method of \citet{Jing2004}:
\begin{align*}
P_s(k, \mu) = 2 \pi \sum_{i,j} & \Delta {r^2_\|}_ i \Delta \ln {r_\bot}_ j
\xi({r_\bot}_j, {r_\|}_i) \cos(k_\| {r_\|}_i) \\
& J_0(k_\bot {r_\bot}_j) W_g({r_\bot}_j, {r_\|}_i),
\end{align*}
where $J_0$ is the zeroth-order Bessel function.
Following \citet{Li2006b}, $r_{\| i}$ runs from $-40$ to 40 \hMpc\ with
$\Delta r_{\| i} = 1 \hMpc$ and $r_{\bot j}$ runs from 0.1 to 50 \hMpc\ with
$\Delta \ln r_{\bot j} = 0.23$.
$W_g$ is a Gaussian window function used to down-weight noisy $\xi(r_\bot, r_\|)$
measurements at large scales:
\begin{equation} \label{eqn:gsmooth}
W_g(r_{\bot}, r_{\|}) = \exp\left(-\frac{r_{\bot}^2 + r_{\|}^2}{2 S^2}\right),
\end{equation}
with smoothing scale $S = 20$ or 25 \hMpc.

As advised by \citet{Jing2001}, we reduce the effects of finite bin sizes in
$r_\|$ and $r_\bot$ by dividing each $r_\|$ and $\ln r_\bot$ bin into 
$N$ sub-bins and interpolate 
$\xi(r_\bot, r_\|)$ at each sub-bin using a bilinear cubic spline.
We have found that $N = 21$ sub-bins is sufficient to obtain reliable
$P_s(k, \mu)$ measurements on small scales (large $k$ values).

\begin{figure}
\includegraphics[width=\linewidth]{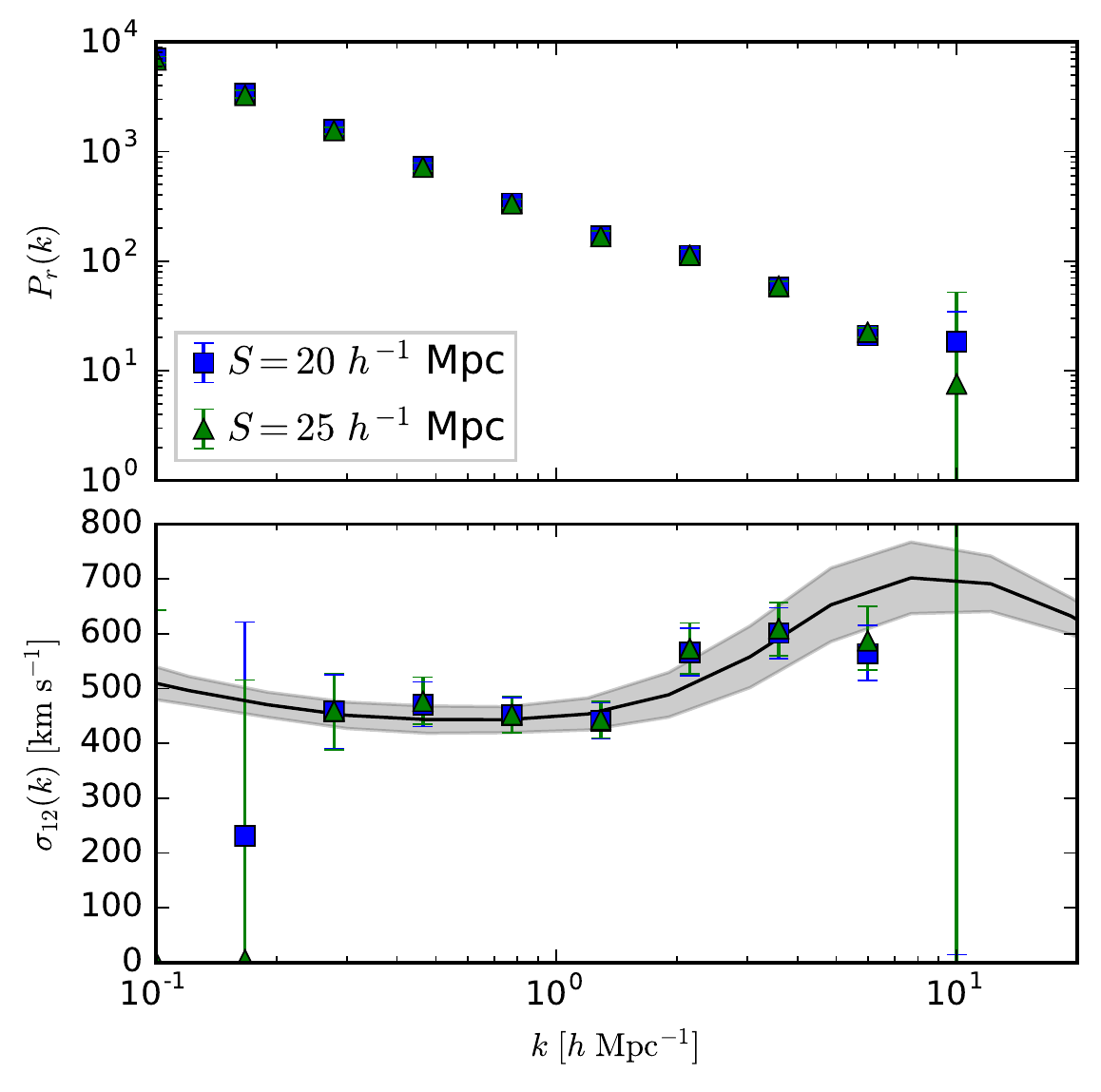}
\caption{Real-space power spectrum $P_r(k)$ (top panel) and pairwise
velocity dispersion $\sigma_{12}(k)$ (bottom panel) estimated from the 
mock catalogues using the methods described in Section~\ref{sec:pvdf}.
The continuous line and shaded region in the lower panel 
show the direct estimate of the real-space exponential velocity dispersion
$\sigma_{12}^{\rm exp}(r)$ and its standard deviation
reproduced from Fig.~\ref{fig:pvd_mock}, assuming that $k = 2\pi/r$.
Blue squares and green triangles respectively denote smoothing scales of 
$S = 20, 25 \hMpc$ in equation~(\ref{eqn:gsmooth}).
}
\label{fig:pvd_k_mock}
\end{figure}

In Fig.~\ref{fig:pvd_k_mock} we show the real-space power spectrum 
$P_r(k)$ and pairwise velocity dispersion $\sigma_{12}(k)$ estimated from the 
mock catalogues by
fitting the model $P_s(k, \mu)$ (equation~\ref{eqn:p_mod}) to the observed one,
assuming a fixed value of $\beta = 0.45$.
The $2 \times n_k$-parameters, 
where $n_k$ is the number of bins in which $P_r(k)$ 
and $\sigma_{12}(k)$ are estimated, and the covariances between the parameters, 
are determined using the {\sc emcee} \citep{Foreman-Mackey2013}
Markov Chain Monte Carlo code.
Both power spectrum and velocity dispersion measurements become very
noisy at small scales, $k \ga 8 \hMpcinv$ or $r \la 0.6 \hMpc$.
At larger scales, $k \la 8 \hMpcinv$,
the PVD estimated in Fourier is in very good agreement with the direct estimate,
particularly with a smoothing length of 25 \hMpc.

\subsection{Summary of tests}

From these tests using mock catalogues, we conclude that the configuration space
dispersion model
provides the most reliable estimate of the PVD on scales 
($0.01 \hMpc\ \la r_\bot \la 10 \hMpc$).
The Fourier-space dispersion model provides reliable estimates
on larger scales ($0.6 \hMpc\ \la r_\bot \ga 30 \hMpc$).
We show results using both of these methods in the following section.

\section{Results and discussion} \label{sec:results}

\begin{figure*}
\includegraphics[width=0.7\linewidth]{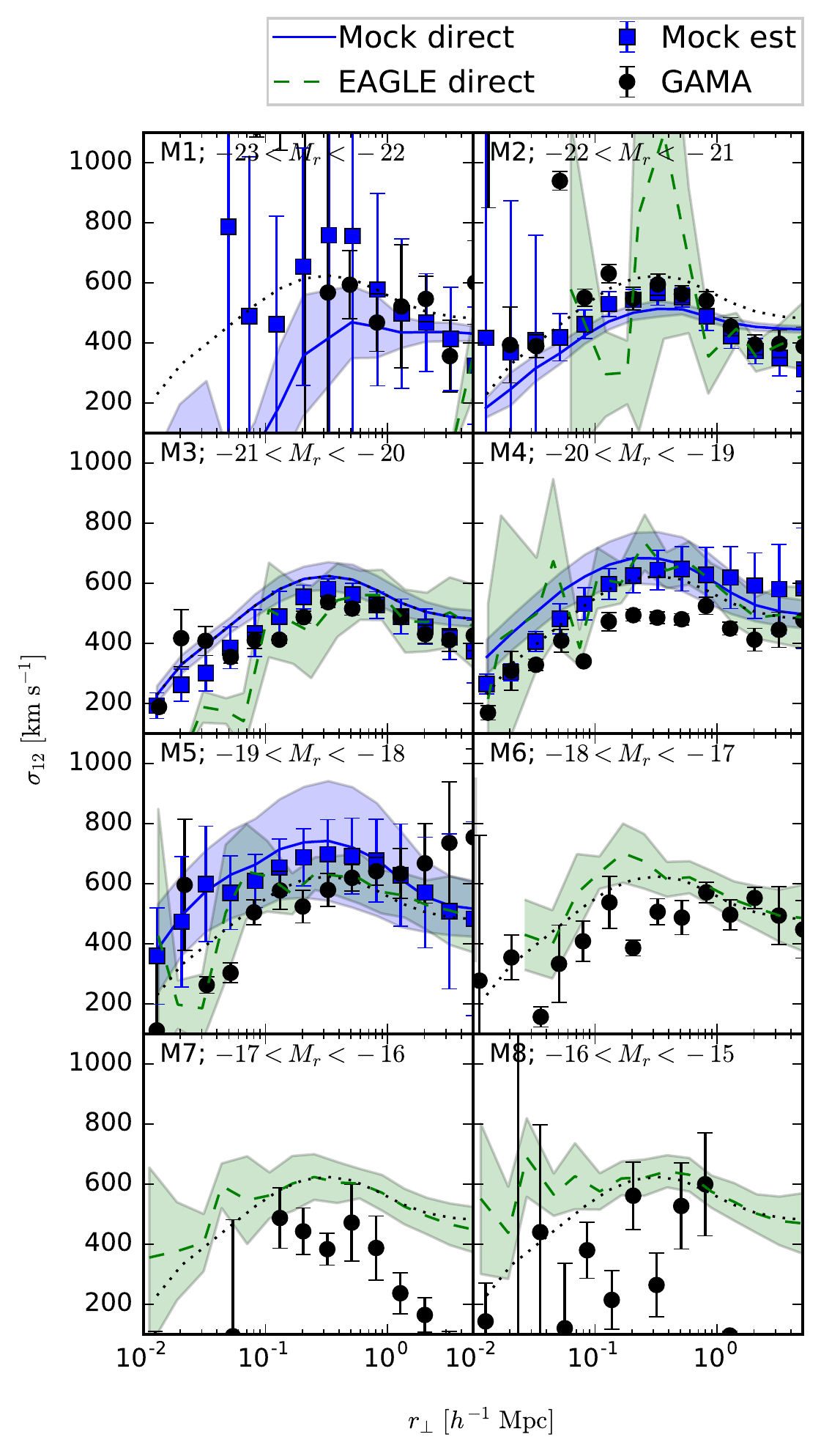}
\caption{Pairwise velocity dispersions with projected separation $r_\bot$
in bins of absolute magnitude as defined in Table~\ref{tab:samples}.
Blue and green lines show direct estimates from the {\sc galform} mocks
and EAGLE simulation respectively,
with the $1\sigma$ error band shaded in both cases.
Blue squares and black circles show dispersion model estimates from the 
mocks and GAMA galaxies respectively.
To aid visual comparison, 
the dotted line reproduces the direct mock estimate for the
M3 sample in the other panels.
The mock error bars show the standard deviation from 26 realisations;
the GAMA error bars are determined from jackknife sampling.
The mock error bars are slightly larger than the GAMA ones,
accounting as they do more fully for sample variance.
}
\label{fig:pvd_beta_lum}
\end{figure*}

\begin{figure}
\includegraphics[width=\linewidth]{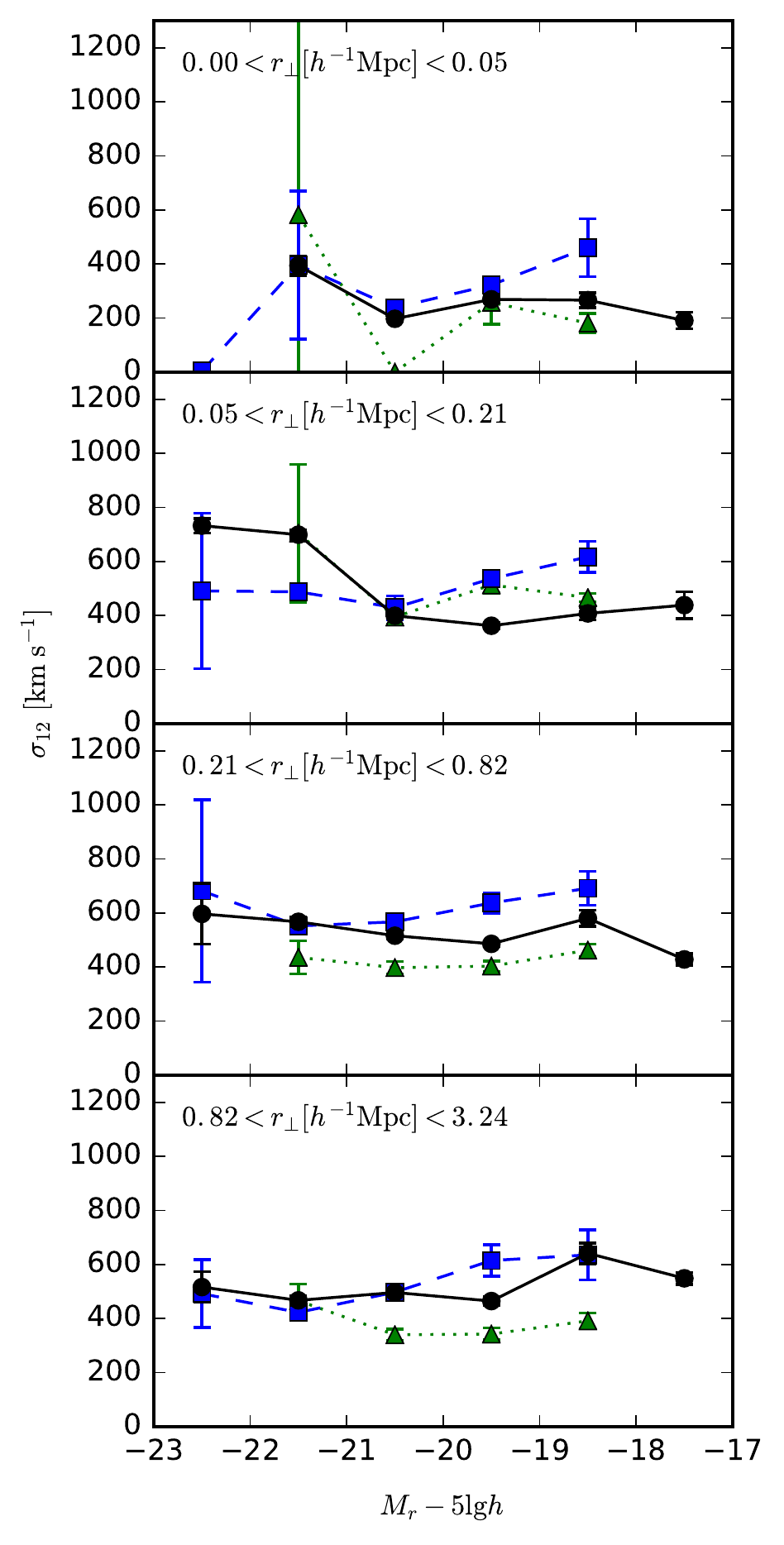}
\caption{Pairwise velocity dispersion for GAMA galaxies 
(black circles connected by continuous lines)
and mocks (blue squares connected by dashed lines) 
as a function of absolute magnitude in bins of
projected separation $r_\bot$ as labelled.
Green triangles connected by dotted lines show the PVD for galaxies 
selected from a volume-limited sample with $M_r < -18$ mag.
}
\label{fig:pvd_lum}
\end{figure}

In Fig.~\ref{fig:pvd_beta_lum} we show the PVD \pvd\ as a function of projected 
separation in bins of absolute magnitude for both GAMA and mock galaxies.
In each panel, for luminosity bins M1--M5,
one should first compare the direct (blue line) and dispersion model
(blue squares) estimates from the mocks.
If these are in good agreement, then the GAMA
dispersion-model results are likely to be reliable.
The space density of galaxies in bin M1 is too low for a reliable estimate
of the PVD on small scales, but the estimates converge for
separations $r_\bot \ga 1 \hMpc$.
For the remaining luminosity bins, M2--M5,
agreement between direct and dispersion-model
estimates of the mock PVDs is excellent on all scales measured.

Comparing GAMA and mock PVDs, for luminous galaxies, bins M1--M3,
the amplitudes are consistent, and both peak at $r_\bot \approx 0.3 \hMpc$.
For fainter galaxies, bins M4 and M5, the mock PVDs are
systematically higher than the GAMA PVDs.
This is particularly noticeable for bin M4,
where the GAMA estimates are unusually low.
Since the large-scale GAMA PVD increases again in lower-luminosity bins,
M5 and M6, this is most likely a sampling fluctuation,
perhaps due to the significant underdensity in the GAMA redshift distribution
around $z \approx 0.2$--0.26, a region from which many of the galaxies in
bin M4 lie \citep[see figs. 4 \& 5 of][]{Farrow2015}.

Since the {\sc galform} mock catalogues are reliable at only relatively
bright magnitudes ($M_r \la -17$ mag), we also compare our
GAMA PVD estimates with those from EAGLE hydrodynamical
simulation \verb|RefL0100N1504|
\citep{Crain2015,Schaye2015,McAlpine2016}.
Placing the observer at the origin of the $z = 0.1$ data cube,
we use the Cartesian velocities of each subhalo to calculate
the line of sight peculiar velocity of each galaxy, and hence the PVD,
as a function of projected separation.
Uncertainties are estimated by subdividing the simulation cube into eight
sub-cubes and calculating jackknife errors.
We have verified that EAGLE $r$-band absolute magnitudes are
consistent with GAMA: the luminosity functions agree extremely well
over the magnitude range $-22 < M_r < -15$.
We therefore use the same absolute magnitude limits when comparing
EAGLE with GAMA.

EAGLE simulation results, shown in Fig.~\ref{fig:pvd_beta_lum} as a green line,
are noisy for luminous galaxies due to the limited volume probed
($10^6$ Mpc$^3$ for $h = 0.6777$).
For moderate luminosities (bins M3--M5),
the agreement with {\sc galform} is good.
The GAMA PVD is also consistent with EAGLE for bin M6, but falls below the
EAGLE prediction for the two faintest bins, M7 and M8.
The GAMA jackknife errors likely underestimate the uncertainties in these
very small volume samples, and so this is not necessarily
indicating a discrepancy with EAGLE at low luminosities.

To show the PVD dependence on luminosity more clearly,
in Fig.~\ref{fig:pvd_lum} we show the PVD \pvd\ as a function of 
absolute magnitude in broad bins of projected separation.
To do this, we determine the average PVD for four sets of three adjacent
separation bins, with separation limits as given in the figure legend.
When averaging, we weight each bin by its inverse-variance,
and the variance on the average is determined in the usual way as the reciprocal
of the sum of inverse variances.

For small scales, $r_\bot \la 1 \hMpc$, corresponding to the top three panels,
the PVD for GAMA galaxies tends to decline near-monotonically from
bright to faint luminosities.
The mock PVD is much flatter, possibly even showing a small increase to
fainter luminosities.
Thus the mocks do a good job at matching the observed PVD for luminous galaxies,
but over-predict the PVD for fainter objects.
The same result was found by \citet{Li2007} when comparing two previous
Millennium-based semi-analytic models \citep{Kang2005,Croton2006}
with SDSS PVD measurements.
\citet{Li2007} show that the mocks most likely place too many faint galaxies
in massive halos.
This problem thus appears to persist in more recent semi-analytic models.
This interpretation is reinforced by the fact that the same mock catalogue 
significantly over-predicts the small-scale projected correlation function 
of faint ($M_r > -18$ mag) galaxies \citep[fig.~11]{Farrow2015}.

One should, however, be aware that the faintest galaxies can only be seen in the
very nearby Universe.
Thus if the local volume is underdense, as has been claimed by several authors,
\citep[e.g.][]{Busswell2004,Keenan2013,Whitbourn2016}, a paucity of local, 
large structures might explain the low observed PVD and projected clustering
for the faintest galaxies.
In order to address this concern, we have defined a second, fainter,
volume-limited sample from GAMA with $^{0.1}M_r < -18$ mag and $z < 0.116$.
We then extract subsamples in bins of absolute magnitude 
$[-22, -21], [-21, -20], [-20, -19], [-19, -18]$.
The galaxies in these absolute magnitude bins are visible throughout the volume
and hence will not suffer from sampling fluctuations due to Malmquist bias.
We show the PVDs from this volume-limited
sample as green triangles in Fig.~\ref{fig:pvd_lum}.
While the trend with luminosity is less clear than for the full GAMA sample,
the results are broadly consistent.
Unfortunately, one cannot extend this analysis to the full luminosity range
plotted due to the tiny volume within which fainter galaxies can be seen.
Note that forming a (separate) volume-limited sample for each luminosity bin
would not alleviate sampling fluctuations.

\begin{figure*}
\includegraphics[width=0.7\linewidth]{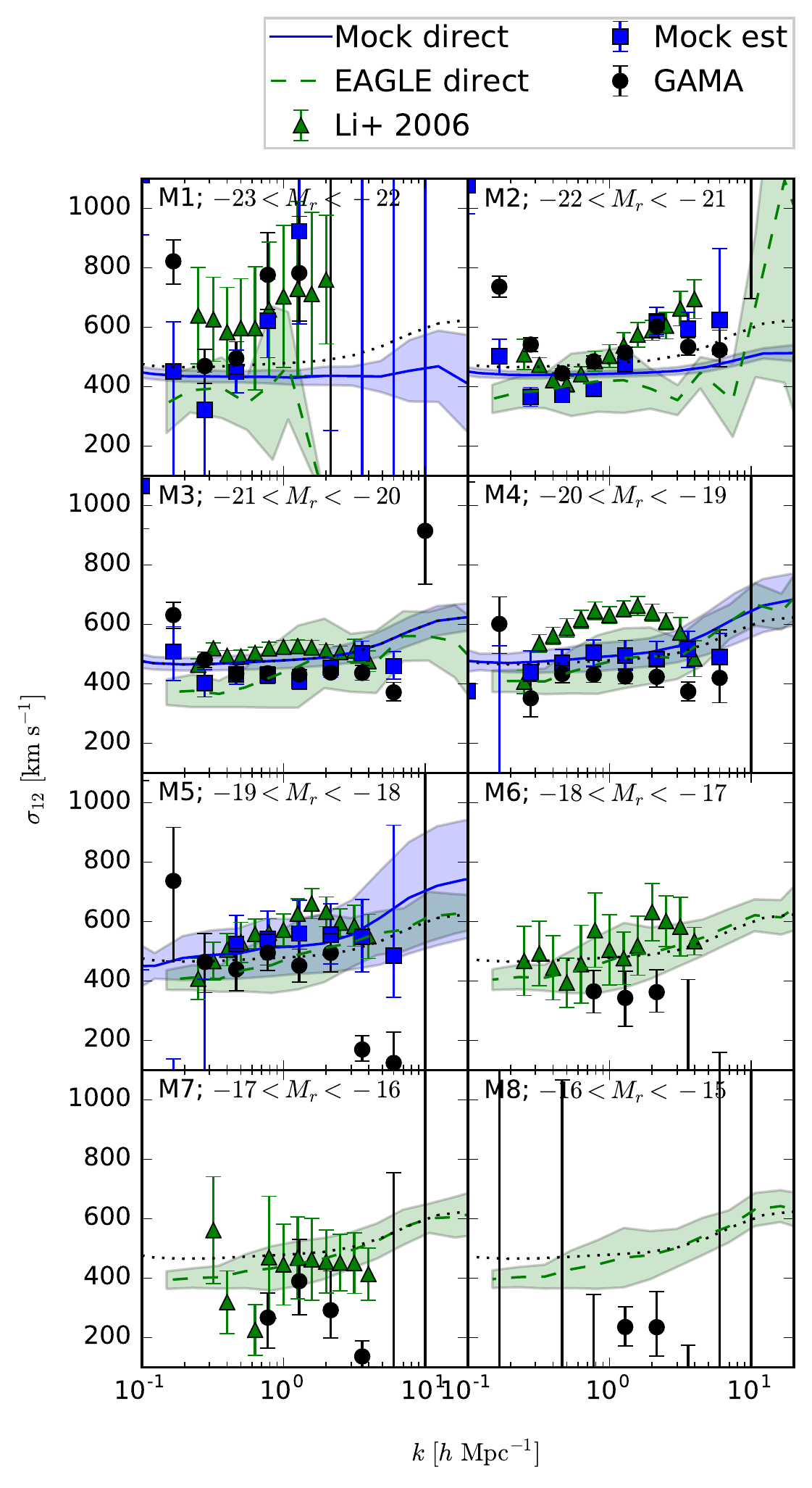}
\caption{Pairwise velocity dispersions with wavenumber $k$
in bins of absolute magnitude.
Blue and green lines show direct estimates from the {\sc galform} mocks
and EAGLE simulation respectively,
with the $1\sigma$ error band shaded in both cases.
Blue squares and black circles show Fourier-space estimates from the 
mocks and GAMA galaxies respectively.
Green triangles show estimates from SDSS \citep{Li2006b}.
To aid visual comparison, 
the dotted line reproduces the direct mock estimate for the
M3 sample in the other panels.
}
\label{fig:pvd_k_lum}
\end{figure*}

In Fig.~\ref{fig:pvd_k_lum} we show the Fourier-based PVD estimate 
as a function of wavenumber $k$.
From comparison with the direct mock estimates, we conclude that these
Fourier-based PVD estimates should be reliable for moderate luminosity galaxies
(bins M2--M5, $-22 \la M_r \la -18$)
over the range of scales $0.2 \la k \la 5 \hMpcinv$.
We also show comparison results from \citet{Li2006b}.
Given the uncertainties, our results are broadly consistent with those of
Li et al. where we overlap, although we do measure systematically lower PVD
than Li et al. for the M4 ($-20 < M_r < -19$ mag) bin.

In future, we plan to 
investigate the dependence of the PVD on location within the cosmic web,
on stellar mass and on redshift.
We also plan to investigate improvements to the mock catalogues to better
match the luminosity-dependence of the observed galaxy PVD,
along with other observational constraints.
In the longer term, it is be hoped that such measurements may be instrumental
in ruling out certain models of modified gravity.

\section{Conclusions} \label{sec:concs}

We have presented measurements of the PVD for luminosity-selected
samples of galaxies from the GAMA equatorial regions, using mock catalogues
to check our estimators.
GAMA's relatively deep flux limit, $r < 19.8$, and high redshift success rate,
$> 98$ per cent, have enabled us to measure the PVD down to a factor of ten
smaller in
projected separation than was possible using SDSS data \citep{Li2007}.
Our findings can be summarized as follows.
\begin{enumerate}
\item In agreement with previous work, \citep[e.g.][]{Hawkins2003}
we find that the form of the pairwise velocity distribution is much better fit
by an exponential than a Gaussian function.
\item The dispersion model can make reliable predictions
of the PVD in configuration space for galaxy pairs with projected separation
0.01--10 \hMpc,
thus allowing detailed tests of galaxy formation models and
hydrodynamical simulations.
\item In Fourier space, one can reliably measure the PVD of GAMA galaxies
for wave numbers in the range 0.2--8 \hMpcinv.
This is similar to the range of scales probed by \citet{Li2007} using SDSS data;
thus the Fourier method employed here does not enable us to exploit the
small-scale fidelity of the GAMA data as well as configuration-space methods.
\item For most luminosity bins, the PVD peaks at $\pvd \approx 600 \kms$ 
at projected separations $r_\bot \approx 0.3 \hMpc$, 
although some fainter bins show a monotonic increase in $\pvd$ with separation.
\item On small scales, $r_\bot \la 1 \hMpc$, 
the measured PVD for GAMA galaxies declines slightly from $\approx 600 \kms$
at high luminosities to $\approx 400 \kms$ at low luminosities.
This trend is not seen at larger scales (0.8--3.3 \hMpc).
\item While the {\sc galform} mocks analysed here give a similar-amplitude
PVD as the GAMA galaxies, they show very little trend with luminosity:
if anything, they predict a slightly increaseing PVD with decreasing luminosity
for $L^*$ and fainter galaxies.
Thus the mocks do a good job at matching the observed PVD for luminous galaxies,
but over-predict the PVD for fainter objects.
\end{enumerate}

\section*{Acknowledgements}

JL acknowledges support from the Science and Technology Facilities Council
(grant number ST/I000976/1).
PN acknowledges the support of the Royal Society through the award
of a University Research Fellowship, the European Research
Council, through receipt of a Starting Grant (DEGAS-259586)
and the Science and Technology Facilities Council (ST/L00075X/1).
We thank the anonymous referee for their insightful comments leading
to clearer presentation.

GAMA is a joint European-Australasian project based around a
spectroscopic campaign using the Anglo-Australian Telescope. The GAMA
input catalogue is based on data taken from the Sloan Digital Sky
Survey and the UKIRT Infrared Deep Sky Survey. Complementary imaging
of the GAMA regions is being obtained by a number of independent
survey programs including GALEX MIS, VST KIDS, VISTA VIKING, WISE,
Herschel-ATLAS, GMRT and ASKAP providing UV to radio coverage. GAMA is
funded by the STFC (UK), the ARC (Australia), the AAO, and the
participating institutions. The GAMA website is:
\url{http://www.gama-survey.org/}.

\bibliographystyle{mnras}
\bibliography{library}

\appendix
\section{Comparison with previous GAMA clustering measurements}
\label{sec:farrow_comp}

\begin{figure}
\includegraphics[width=\linewidth]{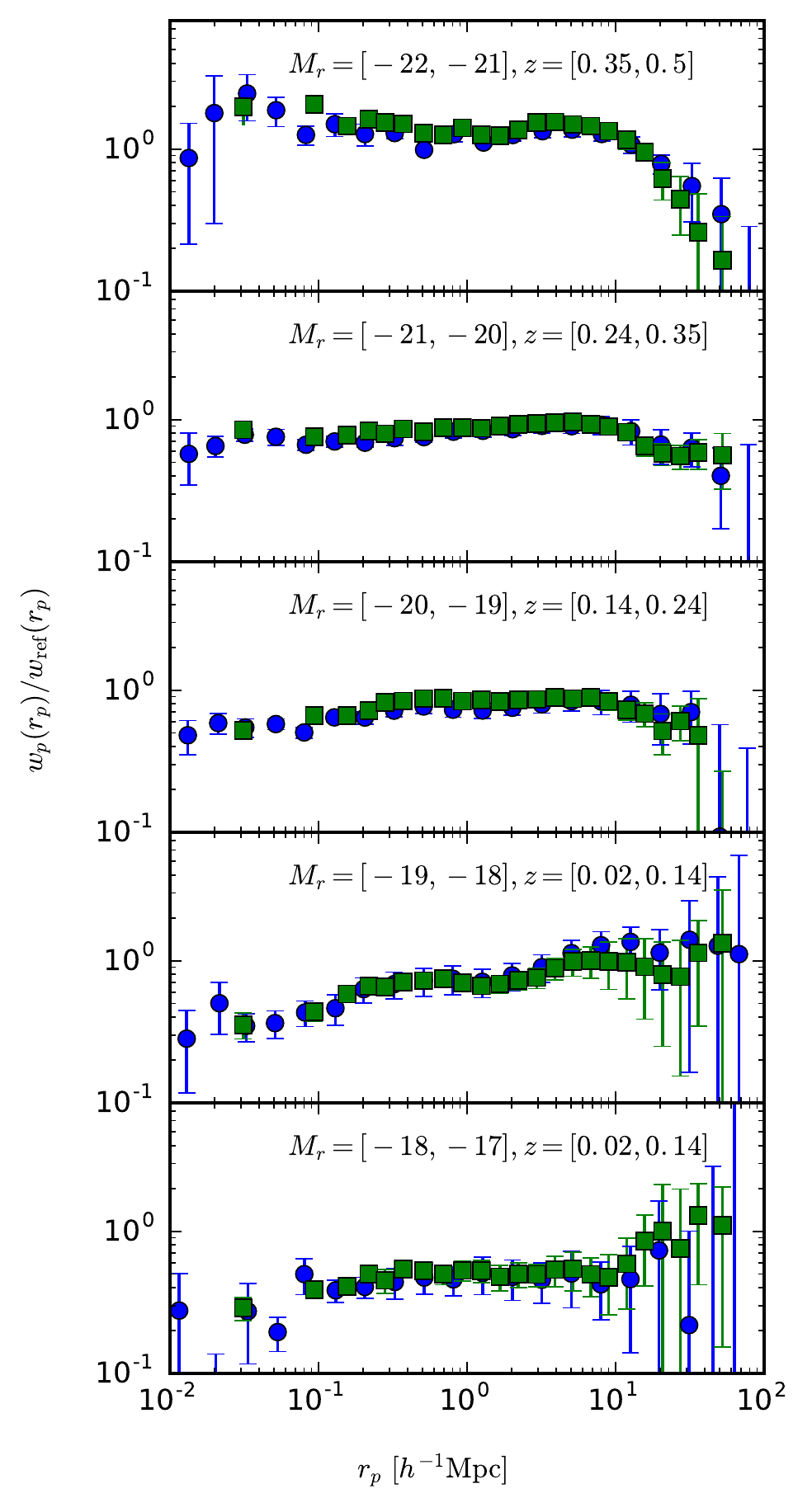}
\caption{Comparison of the projected correlation function for various 
sub-samples of GAMA galaxies.
Blue circles show measurements from the present work; 
green squares show the comparison from \citet{Farrow2015}.
Both sets of measurements have been divided by the reference power-law
used by Farrow et al. in their fig. 10 
(viz $r_0 = 5.33 \hMpc$, $\gamma = 1.81$).
}
\label{fig:farrow_comp}
\end{figure}

A comparison of the projected correlation function for various 
sub-samples of GAMA galaxies from the present work and from \citet{Farrow2015}
is presented in Fig.\ref{fig:farrow_comp}.
We select our samples using the same redshift and luminosity limits as
specified for the largest samples in five $^{0.0}M_r$ absolute magnitude bins
from the top of Table~2 in Farrow et al.
Agreement is excellent on all scales.
We thus find that our independently-determined survey mask and radial selection function 
do not impact measurements of the projected correlation function, on which our 
PVD estimates are based.
%

\section{Accuracy of projected and real-space correlation function estimates} 
\label{sec:xi_test}

\begin{table}
\caption{Testing estimates of $w_p(r_\bot)$ and $\xi_r(r)$ using 26 
volume-limited mock catalogues.
The first row gives the power-law parameters obtained directly from the 
direction-averaged correlation function calculated in real space,
i.e. using cosmological redshifts.
The remaining rows show results obtained by integrating $\xi(r_\bot, r_\|)$ 
up to different values of ${r_\|}_{\rm max}$ in equation~(\ref{eq:proj}).
The first column gives the upper integration limit, ${r_\|}_{\rm max}$, 
the second and third columns the mean and standard deviation of the 
recovered power-law parameters $\gamma$ and $r_0$.
The fourth column ($\chi^2_d$) gives the $\chi^2$ residual 
(for 20 degrees of freedom)
between the direct estimate of $\xi_r(r)$ using cosmological redshifts, 
and the non-parametric estimates obtained by inverting $w_p(r_\bot)$.
The fifth column ($\chi^2_{\rm pl}$) gives the $\chi^2$ residual 
(for 11 degrees of freedom) for the power-law fit to the 
non-parametric estimate.
}
 \label{tab:rmax-test}
 \begin{math}
 \begin{array}{rrrrr}
 \hline
\multicolumn{1}{c}{{r_\|}_{\rm max}\ [\hMpc]} & \multicolumn{1}{c}{\gamma} & 
\multicolumn{1}{c}{r_0\ [\hMpc]} & \multicolumn{1}{c}{\chi^2_d} & 
\multicolumn{1}{c}{\chi^2_{\rm pl}} \\
 \hline
{\rm Direct} & 1.84 \pm 0.01 & 4.84 \pm 0.24 & & 60\\
 \hline
10 & 1.86 \pm 0.01 & 5.09 \pm 0.23 & 141 & 22\\
20 & 1.85 \pm 0.02 & 4.95 \pm 0.31 & 20 & 29\\
30 & 1.86 \pm 0.02 & 4.82 \pm 0.33 & 14 & 27\\
40 & 1.87 \pm 0.02 & 4.85 \pm 0.32 & 10 & 20\\
50 & 1.87 \pm 0.02 & 4.85 \pm 0.34 & 11 & 17\\
60 & 1.86 \pm 0.03 & 4.86 \pm 0.37 & 12 & 17\\
100 & 1.88 \pm 0.03 & 4.67 \pm 0.39 & 12 & 15\\
 \hline
 \end{array}
 \end{math}
\end{table}

\begin{figure}
\includegraphics[width=\linewidth]{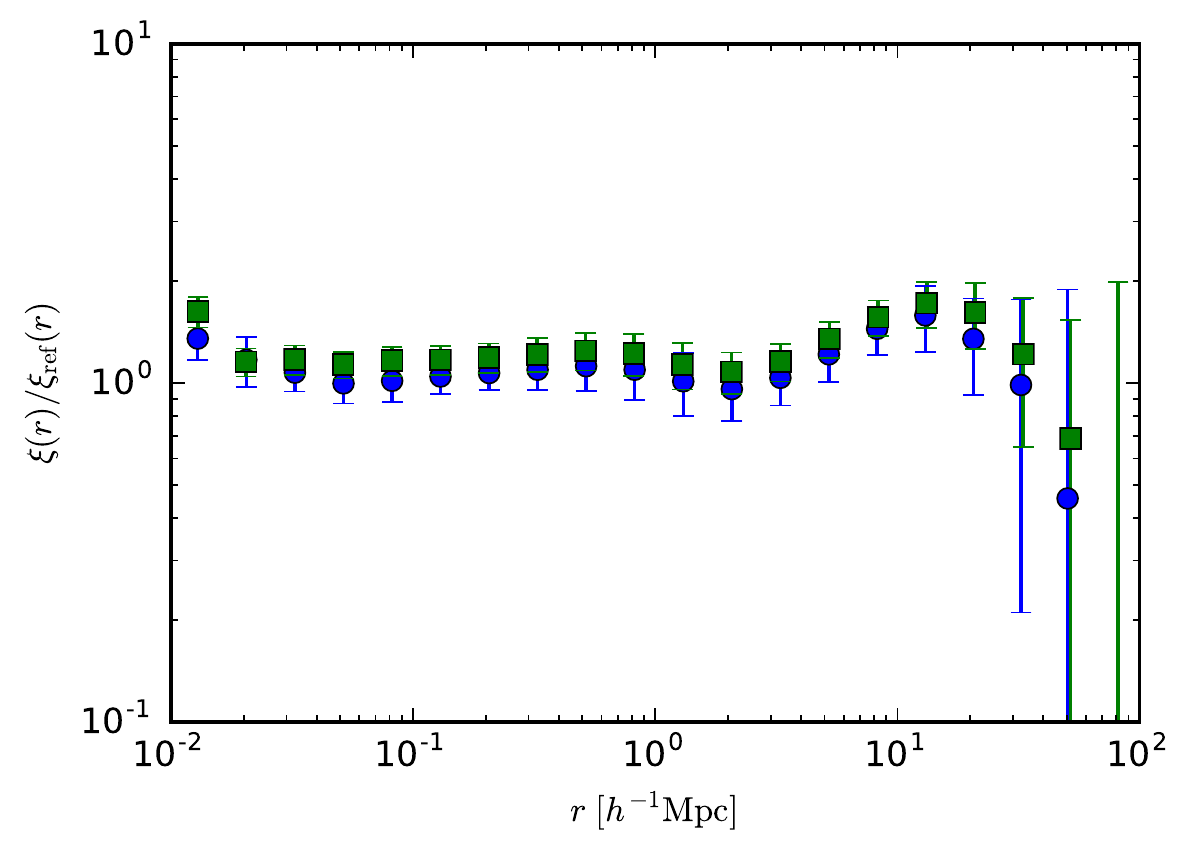}
\caption{Average real-space correlation function $\xi_r(r)$ measured from 
26 mock catalogues
using direction-averaged correlation function and cosmological redshifts
(green squares) and estimated from projecting and inverting the two-dimensional
correlation function (equations \ref{eq:proj} and \ref{eq:real} with
${r_\|}_{\rm max} = 40 \hMpc$, blue circles).
Both sets of measurements have been divided by the power-law fit to the
direct $\xi_r(r)$ measurement
(viz $r_0 = 4.84 \hMpc$, $\gamma = 1.84$).
}
\label{fig:xir_mock}
\end{figure}

We here test how accurately one can recover the projected, $w_p(r_\bot)$,
and real space, $\xi_r(r)$, correlation functions from the two-dimensional 
redshift-space correlation function $\xi(r_\bot, r_\|)$, making use of the 
GAMA mock catalogues.
Since these mock catalogues include cosmological redshift,
due purely to Hubble expansion, as well as
observed redshift, including the LOS component of peculiar velocity, 
one can obtain a direct estimate of $\xi_r(r)$ using equation~(\ref{eq:ls}), 
with galaxy coordinates determined using cosmological 
redshift and counting galaxy and random pairs as a function of total separation.

Carrying out this procedure for each of our mock catalogues, 
we obtain a real space correlation function well-described by a power law
$\xi_r(r) = (r/r_0)^{-\gamma}$ over the range of separations 0.01--5 \hMpc\
with parameters given in Table~\ref{tab:rmax-test}.

We then calculate the projected correlation function $w_p(r_\bot)$
for each of the mock catalogues as described in Section~\ref{sec:xi2d},
fit a power-law over the same range of scales, 
and use equation~(\ref{eq:power-law})
to find the real-space power-law parameters $r_0$ and $\gamma$.
We next invert $w_p(r_\bot)$ using equation~(\ref{eq:real}) to obtain $\xi_r(r)$
estimates, and calculate the $\chi^2$ residuals from the direct estimate.
For both power-law fits and $\chi^2$ estimates, we utilise 
the full covariance matrices of $w_p(r_\bot)$ and $\xi_r(r)$, respectively.

Our results, obtained using different values of ${r_\|}_{\rm max}$ in 
equation~(\ref{eq:proj}), are given in Table~\ref{tab:rmax-test}.
These results show that the $\chi^2$ residual between direct
and indirect estimates of $\xi_r(r)$ is minimised for 
${r_\|}_{\rm max} = 40 \hMpc$.
Moreover, power-law fits to $w_p(r_\bot)$ have converged by this point;
therefore we use ${r_\|}_{\rm max} = 40 \hMpc$ when calculating
$w_p(r_\bot)$ from the GAMA data.

In Fig.~\ref{fig:xir_mock} we compare the real-space correlation function
from the mock catalogues obtained directly using the cosmological redshifts,
and using equations \ref{eq:proj} and \ref{eq:real} with
${r_\|}_{\rm max} = 40 \hMpc$.
The deprojected correlation function is systematically lower than the directly
measured one, but the bias is within the standard deviation of each measurement.

\section{Hamilton linear infall equations}
\label{sec:hamilton}

\citet{Kaiser1987} showed that coherent infall in Fourier space leads
to a redshift-space power spectrum $P_s(k) = (1 + \beta \mu_k^2) P_r(k)$.
\citet{Hamilton1992} translated this into configuration space to show that
the redshift-space correlation function is given by
\begin{equation}
  \xi'(r_\bot, r_\|) = \xi_0(s) P_0(\mu) + \xi_2(s) P_2(\mu) +
  \xi_4(s) P_4(\mu),
\end{equation}
where the $P_l(\mu)$ are Legendre polynomials.
The harmonics of the correlation function are given by
\begin{align}
  \xi_0(s) &= \left( 1 + \frac{2 \beta}{3} + \frac{\beta^2}{5} \right) \xi(r),
  \label{eqn:xi0} \\
  \xi_2(s) &= \left( \frac{4 \beta}{3} + \frac{4 \beta^2}{7} \right)
  [\xi(r) - \overline{\xi}(r)],\\
  \xi_4(s) &= \frac{8 \beta^2}{35} \left[ \xi(r) + \frac{5}{2} \overline{\xi}(r) -
             \frac{7}{2} \overline{\overline{\xi}}(r)\right], \label{eqn:xi4}
\end{align}
where
\begin{align}
  \overline{\xi}(r) &= \frac{3}{r^3} \int_0^r \xi(r') r'^2 dr', \\
  \overline{\overline{\xi}}(r) &= \frac{5}{r^5} \int_0^r \xi(r') r'^4 dr'.
\end{align}
For a power-law form for the correlation function,
$\xi(r) = (r/r_0)^{-\gamma}$,
equations (\ref{eqn:xi0}--\ref{eqn:xi4}) reduce to \citep{Hawkins2003}
\begin{align}
  \xi_0(s) &= \left( 1 + \frac{2 \beta}{3} + \frac{\beta^2}{5} \right) \xi(r),\\
  \xi_2(s) &= \left( \frac{4 \beta}{3} + \frac{4 \beta^2}{7} \right)
  \left( \frac{\gamma}{\gamma - 3} \right) \xi(r), \\
  \xi_4(s) &= \frac{8 \beta^2}{35} \left[ \frac{\gamma (2 + \gamma)}{(3 - \gamma)(5 - \gamma)} \right) \xi(r).
\end{align}

\bsp	
\label{lastpage}
\end{document}